# Comparison the optical properties for $Bi_2O_3$ and NiO ultrathin films deposited on different substrates by DC sputtering technique for transparent electronics.

M. RASHEED, R. BARILLÉ
MOLTECH-Anjou, Université d'Angers/UMR CNRS 6200,
2 Bd Lavoisier, 49045 Angers, (France)
Corresponding author: rasheed.mohammed40@yahoo.com

## Abstract

Bismuth and Nickel transparent oxides thin films were grown on glass and flexible polythelene terephalate (PET) substrates by DC sputtering technique at room temperature 300 K. The structures of $Bi_2O_3$ and NiO films were analyzed by X-ray diffraction (XRD) analyses and scanning electron microscopy (SEM). A profilometry is used to measure the thicknesses of the ultrathin films. These values were established by a spectro-ellipsometry technique with a new amorphous dispersion model in the range of 200 nm to 2200 nm with one nm increment. In addition, the optical constants of these films were investigated using a double beam UV-Vis-NIR spectrometer and a spectro-ellipsometry and compared using the same wavelength range. They present an excellent agreement. According to the optical transmittance spectra of the films a high average transmittance in the visible region of each sample deposited onto different substrates is measured. Moreover, the absorption coefficients and the optical energy gaps for four types of optical transitions (allowed direct, forbidden direct, allowed indirect, forbidden indirect) of the films were determined and compared with the two different techniques. Finally, the nature of the structure and morphology of $Bi_2O_3$ and NiO ultrathin films has been analyzed and compared.

## Keywords

Bismuth and nickel oxides films, DC sputtering, different substrates, spectrophotometer, ellipsometry.

## 1. Introduction

Bismuth oxide ($Bi_2O_3$) and nickel oxide (NiO), (metal oxides thin films) are P-type semiconductor [1, 2], antiferromagnetic [3, 4], and are interesting for their excellent optical and electrical properties, high chemical thermal stability, low resistivity, high transparency in the visible region, and a wide optical energy gap [5, 6]. These films exhibit an extremely wide range of physical and chemical properties used in a variety of technical applications that includes the uses in photoelectrochemical or photocatalysis [7, 8], smart windows [9, 10], electrochromic devices [11, 12], transparent electronic devices [13, 14], gas sensing [15], metal-insulator [16, 17], fuel cells [18, 19], and heat mirror [20]. The surface and interface properties of metal oxides play an important role in all these applications, and sometimes even dominate device performances. Among them Bismuth oxide and Nickel oxide play an important role due to their high melting points (817°C, 1955°C) respectively remaining until this reached temperature point with a high corrosion resistance. $Bi_2O_3$ and NiO are also good candidates as electrodes for thin film solar cell applications because of high thermal expansion coefficients representing large dimensional variations under heating and cooling, which would limit the performance of an electrolyte. NiO (octahedral) is commonly known as the rock salt structure, and is often non-stochiometric. The stoichiometric NiO is green and the non-stoichiometric NiO is black. There are various techniques for the growth of these films such as molecular beam epitaxy [21, 22], chemical vapor deposition (CVD), thermal evaporation [23, 24], spray pyrolysis [25, 26], sol gel [27, 28], pulsed laser deposition [29, 30], RF reactive sputtering [31, 32], induced self-assembly [33, 34], DC sputtering [35, 36]. However, DC sputtering has become an attractive method for nanomaterials. In this way, the increasing of substrate temperatures was not used, which is very important with flexible Polyethylene terephthalate (PET) substrate to prevent it from damages.

Until now, we did not find any studies for the $Bi_2O_3$ and NiO thin films prepared by DC sputtering method and deposited on different substrate like glass and PET. The optical properties of these thin films were investigated using spectrophotometry and spectroscopic ellipsometry methods with a comparison between them. The aim of the present work is to synthesis and studies the optical constants of $Bi_2O_3$ and NiO ultrathin films deposited on different substrates at room temperature by a DC sputtering method. The optical properties of the deposited films were determined using spectrophotometry (UV) and spectroscopic ellipsometry (SE) in the spectral range 200 - 2200 nm. The objective of the present work paper is to investigate the optical properties of $Bi_2O_3$ and NiO thin films that are candidate for future of transparent conducting oxide (TCOs). The DC sputtering technique was used to prepare the films in the same deposition conditions at 300 K. The optical constants were calculated with a new amorphous model and the dispersion formula has been used for these determinations. As a result, a high transparency in the visible region and a wide optical band gap makes these materials excellent and promising for solar cells, anti-reflecting coatings, electronics, optoelectronic and other applications.

## 2. Experimental

### 2.1 Films fabrication

A DC sputtering technique was used to deposit $Bi_2O_3$ and NiO thin films onto both glass and PET substrates at room temperature. Commercial Corning microscope glass-slides, 75×50 mm from (TED PELLA INC.) were cut into $25x20$ mm plates with a thickness of $1.1\,mm$. Flexible PET substrate (HIFI PMX739, with a size of $25\ x\ 20$ mm and a thickness of $175\ \mu m$) were used as substrates in this work. Glass substrates were cleaned carefully in an ultrasonic bath treatment type (BRANSON Ultrasonic-CAMDA 19 spc) with ethanol, acetone, deionized water and dichloromethane, each one for 20 min and dried with nitrogen gas jet. The distance between the target-substrate was kept at about $7\ cm$, and the deposition time was: $4\ nm/min$ and the pressure was P = $5 \times 10^{-2}\ mb$ and the sputtering power of 100W at 300K for $Bi_2O_3$ and NiO films.

### 2.2 Film thickness measurements

Film thicknesses were measured by a Surface Profilometer with a Veeco 6M Metrology L.L.C giving 20.98 and 20.78 nm for $Bi_2O_3$ ultrathin films deposited on glass and PET substrates respectively. The thicknesses were about 20.43 and 20.55 nm for the NiO thin films deposited on glass and PET. While using SE measurements, the film thicknesses for $Bi_2O_3$ ultrathin films on glass and PET were in the range of 20.00- 20.21 nm and 20.01-20.44 nm for NiO ultrathin films deposited on glass and PET substrates respectively. The average error percentage between the exact and measured values of the film thicknesses was 4%.

### 2.3 Structural properties

The structures of the samples were analyzed by a D8 Advance Brucker diffractometer $CuK\alpha\,1{,}2\ (\lambda = 1.5406\ A^o)$ - (XRD) X-ray diffraction. The crystal structural of the thin films deposited on glass and PET substrates examined by X-ray diffraction using standard $Cuk\alpha$ radiation $(\lambda =$ $0.15406\ nm$) appeared amorphous. As it is known crystalline growth peak of any thin films occurs for the thin films that grown at room temperature, especially for the ones that have a thickness smaller than 100 nm. The thickness parameter is very important for influencing crystallization. The surface morphology of the $Bi_2O_3$ and NiO nanoparticle film on both glass and PET substrates has been examined by scanning electron microscopy (SEM).

### 2.4 Optical properties

The optical properties: transmittance $(T)$, reflectance$(R)$, and absorption $(\alpha)$ spectra were obtained at ambient temperature in the wavelength ranges of 200-2200 nm by a PerkinElmer Lambda 950 (UV/Vis/NIR) Spectrophotometer. In addition, the skin depth spectra $(X)$ and the optical band gap $(E_g)$, were determined. The optical constants were made in the same spectral range of the spectrophotometer using a phase modulated spectroscopic ellipsometry type (UVISEL NIR Horiba Jobin Jvon). The angle of incident is $70^0$ with an increment of one nm. The spectroscopic Ellipsometry (SE) is a well-known non-destructive and a very sensitive optical method for measuring film thicknesses and optical properties of materials by measuring the effect of reflection on the polarization state of light. It is defined by the relationships between the amplitude $\psi$ and phase $\Delta$ angles as a function of wavelengths in terms of angles of incidence of the light reflected from the surface of the film, which is coated on a substrate. The two component plane waves of the electric field is resolved. The first component is called P-wave with the electric field parallel to the plane of incidence and the second component is called S-wave in the case of perpendicular component in the plane of incidence. The resultant wave is a plane polarized if P- and S- components are in phase ($0^0$) or out of phase ($180^0$). If the phase difference $\Delta = \delta_p - \delta_s$ is greater than $180^0$ this case is called elliptical polarized. As a result, reflection causes a change in relative phases of the P and S waves and the ratio of their amplitude changes in phase is represented by an angle $\Delta$ and the amplitude ratio change is represented by an angle $(\psi)$. The ellipsometry measurements are described by the complex reflectance ratio $(\rho)$, which is expressed by the ratio between the parallel $(p -)$ and perpendicular $(s -)$ polarized light (Fresnel reflection coefficients) leading to the electrical field alteration after reflection [37]:

$$\rho = \left|\frac{r_p}{r_s}\right| = tan\psi e^{i\Delta} \qquad 1$$

where $r_p$ and $r_s$ are the Fresnel reflection coefficients for the P- and S-polarized light respectively.

The ellipsometry setup consists in a monochromatic light source, polarizer, analyzer, compensator, and detector. In order to analyze the film surface and to find the optical properties of the film after the SE measurements, the data were modeled with an appropriate dispersion model and the experimental results must be fitted with this model and compared with the theoretical data. Ellipsometry data can be analyzed by Delta Psi2 software from (Jobin-Yvon Co.). The values of $\psi$ angle range are between (0-90$^0$) while $\Delta$ angle range is between (0-360$^0$). The new Amorphous model works very well for amorphous materials exhibiting an absorption in the visible and FUV region and is used as a mathematical model to describe the optical properties of thin films. This model can be expressed with the Equations 2-5 [38-41]:

The extinction coefficient can be expressed by the following equation

$$k(\omega) = \begin{cases} \frac{f_j.(\omega-\omega_g)^2}{(\omega-\omega_j)^2+\Gamma_j^2}, & for\ \omega > \omega_g \\ 0, & for\ \omega \leq \omega_g \end{cases} \qquad 2$$

The refractive index can be expressed by the following equation

$$n(\omega) = \begin{cases} n_\infty + \frac{B_j.(\omega-\omega_j)^2+c}{(\omega-\omega_j)^2-\Gamma_j^2}, & for\ \ \omega > \omega_g \\ 0, & for\ \ \omega \leq \omega_g \end{cases} \qquad 3$$

where

$$B_j = \frac{f_j}{\Gamma_j}\left(\Gamma_j^2 - \left(\omega_j - \omega_g\right)^2\right) \qquad 4$$

$$c_j = 2.f_j.\Gamma_j.\left(\omega - \omega_g\right) \qquad 5$$

where the number of oscillators equal 1 while the number of parameters is equal 5. $n_\infty$ is the refractive index. It is an additional parameter with a value higher than 1 and equal to the value of the refractive index when ($\omega \to \infty$). Four parameters are used to describe the extinction coefficient $k$: 1) $f_j$ is the peak of the extinction coefficient, 2) $\Gamma_j$ is the broadening term of the peak of absorption, 3) $\omega_j$ is the energy at which the extinction coefficient is maximum and 4) $\omega_g$ is the energy of optical band gap $E_g$.

## 3. Results and discussions

### 3.1 Structural and morphological analysis

The crystal structure of the Bismuth trioxide and Nickel oxide ultra-thin films deposited on glass and PET substrates and examined by X-ray diffraction using a standard $Cuk\alpha$ radiation ($\lambda = 0.15406\, nm$) are amorphous. The structural morphology of these thin films deposited on glass and PET substrates has been analyzed by a scanning electron microscope (SEM). As we have observed from the micrography, the nanoparticle films are very thin, uniformly distributed on both substrates and those on a glass substrate have a surface smother than those on PET substrate. The larger grains of the films are absorbed on the PET substrates.

The influence of the substrate nature on the crystallite growth of the thin $Bi_2O_3$ and NiO films deposited on both glass and PET substrates is clearly observed, while the genesis of crystalline grain size on the PET gives larger grains compared to the same film deposited on glass substrates as shown in figure 1.

### 3.2 Optical properties by UV-vis-NIR spectrophotometer

The optical transmission and reflectance spectra of the $Bi_2O_3$ and NiO thin films deposited on glass and PET substrates with the thicknesses of 20 nm were measured in the spectral range between (200-2200 nm) to derive the optical constants of the films and are presented in Fig. 2 (a and b).

It is noticed in the figure 2 that the transmittance varies with wavelengths: decreasing rapidly from a value of about 60% at 371 nm to a minimum value of 20% at 330 nm and from 60% at 375 nm to 20% at 333 nm for $Bi_2O_3$ films on glass and PET substrates respectively. For NiO the transmittance changes from 60% at 380 nm to 20% at 269 nm and from 60% at 334 nm to 20% at 261 nm on glass and PET substrates respectively. From the minimum value the transmittance increases sharply to the maximum values of 87% and 81% at 550nm for $Bi_2O_3$ on the two different substrates and 77% to 73% at 550nm for NiO on the two substrates. The graphs of reflectance as a function of wavelengths for the films are shown in inset of the Figure 1 a and b. The reflectance of $Bi_2O_3$ and NiO films varies with wavelengths in the same manner: increasing sharply at 200 nm to the first maximum values of about 21% to 23% at 350 nm to 366 nm and to the second maximum values of about 13.8% to 14% at 810 nm to 875 nm for $Bi_2O_3$ on glass and PET substrates respectively. The transmittance increases sharply to the first maximum values of about 22% to 27% at 351 nm to 292 nm for NiO films on two substrates and then decreases sharply to minimum values of about 6.4 % to 7% at 2200 nm for $Bi_2O_3$ and 15% to 14.7% at 2200 nm for NiO. However, the decrease of the reflectance of $Bi_2O_3$ is observed to be greater than the one of NiO.

The investigation of optical absorption spectrum is one of the most fruitful tools for understanding and developing the band structure and optical band gap energy ($E_g$) of amorphous or crystalline structure. The figure 3 shows the dependence of absorption coefficient α on the wavelengths in the spectral range 200-2200 nm for $Bi_2O_3$ and NiO thin film on glass and PET. Based on this figure it is clear that the absorption coefficient of the $Bi_2O_3$ and NiO films varies with wavelengths in the same manner: increasing from values of about $8.8x10^4$ $cm^{-1}$ to $1.48x10^5$ $cm^{-1}$ at 400 nm for $Bi_2O_3$ on glass and PET and from $5.3x10^5$ $cm^{-1}$ to $2.7x10^5$ at 300 nm for NiO on glass and PET to various maximum values of $4.8x10^6$ $cm^{-1}$ to $1.4x10^6$ $cm^{-1}$ at 200 nm for $Bi_2O_3$ on glass and PET. Moreover, the absorption coefficient changes from $4x10^6$ $cm^{-1}$ to $5.9x10^6$ $cm^{-1}$ at 200 nm for NiO on glass and PET and thereafter decreases sharply with increasing wavelengths to zero at 2200 nm. The differences of the absorption $\alpha$ of films in the UV range due to the effect of the optical band gap is given in the figure 3. These results confirm that the absorption is in the UV range of these films.

The optical band-gap value could be obtained from the optical absorption spectrum by using the Tauc's formula [43]

$$(\alpha . h\upsilon)^n = B(h\upsilon - E_g) \qquad 6$$

where, $\alpha$ is the absorption coefficient, $(h\upsilon)$ is the photon energy, $B$ is a constant and $n$ suppose the values 1/2, 2, 3/2 and 3 for allowed direct, allowed indirect, forbidden direct and forbidden indirect transitions, respectively. The figure 4 (a, b, c and d) illustrates the variation of $(\alpha . h\upsilon)^n$ versus $h\upsilon$ for these transitions for $Bi_2O_3$ and NiO thin film. The straight-line nature of the plots over a wide range of photon energy indicates the type of transition. Then, the optical gaps have been calculated by the extrapolation of the linear portion on the energy axis as shown in figure 4. The optical gap for $Bi_2O_3$ and NiO films on glass and PET substrates for different electronic transition types is shown in the table 1.

In Table 1 it can be seen that the value of films with direct band gap is greater than those of indirect band gap deposited on two types of substrates. The optical band gap values of $Bi_2O_3$ for thin films is smaller than those of NiO films for direct transition while those values are greater than NiO films for indirect transition.

### 3.2 Calculations of optical constants by means of spectro-ellipsometry

The optical properties of $Bi_2O_3$ and NiO films have been determined using SE measurements by an optical new amorphous model. Ellipsometry does not actually measure optical constants or film thicknesses, but gives the parametrers $\Psi$ and $\Delta$ function of these characteristics. In our case, the measurements were conducted for wavelengths in the range of 200-2200 nm with an increment of 1 nm at an angle of incident ($70^o$) and the thickness of the glass and PET substrates is assumed nearly infinite as compared to those of $Bi_2O_3$ and NiO films. In the model in order to obtain the optical properties and constants of these films, the optical parameter data are fitted with the DeltaPsi2 software. The mean square error $\chi^2$ value is used to appoint the difference between the experimental and theoretical results. This value should be as small as as possible. The $\chi^2$ is defined as follows [44]:

$$\chi^2 = min \sum_1^n \left[ \frac{(\Psi_{th} - \Psi_{exp})_i^2}{\Gamma_{\Psi,i}} + \frac{(\Delta_{th} - \Delta_{exp})_i^2}{\Gamma_{\Delta,i}} \right] \qquad 7$$

where $\Gamma_i$ is the standard deviation of the points. The smallest value of $\chi^2$ refers to a better fitting results.

For the structure of $Bi_2O_3$ and NiO films on glass or PET substrates, the model is a three-layer model. The bottom layer is full of the void layer, while the middle layer is full of the substrate layers (glass or PET) and the upmost layer is the thin film layer $Bi_2O_3$ and NiO.

The fixed index dispersion formula was used for the void layer in order to derive the optical properties of this layer. The assumed model is given by constant refractive and extinction coefficients for any wavelength using the following equation [44]

$$n(\lambda) = constant = n, k(\lambda) = constant = k \qquad 8$$

where $n$ and $k$ is the value of the refractive and extinction indices as a function of wavelengths respectively.

Equation 8 is applicable to non-dispersive materials like "air" and does not apply to metal and semiconductors. In our work, the physical parameters of this layer are $n = 1.5$ and $k = 0$.

New amorphous dispersion formula was used for glass layer. This formula is used in order to give a Lorentzian shape to the expressions of the refractive and extinction coefficients. Equations 2, 3, 4 and 5 in section 2.4 of this paper expressed this dispersion formula. The physical parameters of the glass layer are listed in Table 2.

The Tauc-Lorentz dispersion formula are used to describe the physical parameter of PET layer. Jellison and Modine developed this model [40] using Tauc joint density of states and Lorentz oscillator. This formula is applied to describe the complex dielectric function $\tilde{\varepsilon}_T = \varepsilon_r + i\varepsilon_i$ where $\varepsilon_r$ and $\varepsilon_i$ is the real and imaginary parts of dielectric constant. The number of oscillators are used in Tauc-Lorentz 3 where (N=3), therefore, the number of parameters was 11. The real and imaginary part of Tauc-Lorentz formula are expressed as follows

$$\varepsilon_r(E) = \varepsilon_r(\infty) + \frac{2}{\pi} P \int_{E_g}^{\infty} \frac{\xi.\varepsilon_i(\xi).d\xi}{\xi^2 - E^2} \qquad 9$$

$$\varepsilon_i = \begin{cases} \frac{1}{E} \cdot \frac{A_i . E_i . C_i . (E-E_g)^2}{(E^2-E_i^2)^2 + C_i^2 . E^2}, & for\ E > E_g \\ 0 & , for\ E \leq E_g \end{cases} \quad 10$$

where $P$ is the Cauchy principal value, the $(i)$ refers to the number of oscillators; in our case $i = 3$.

One parameter is linked to the real part of dielectric function, $\varepsilon_r(\infty) = \varepsilon_\infty$ is the high frequency dielectric constant. Four parameters are used to describe imaginary part of the dielectric function. $A_i$ is related to the strength of the absorption peak. $C_i$ is the broadening term. It is a damping coefficient linked to FWHM of the $i^{th}$ peak of absorption. $E_g$ is the Tauc's optical band gap energy. E is the energy of maximum transition probability or the energy position of the peak of absorption.

The physical parameters of the PET layer in our case are listed in Table 3.

For the $Bi_2O_3$ and NiO layer, a new amorphous dispersion function was used for the glass layer. The experimental curves and those obtained by a new amorphous fitting model for the parameters $\Psi(\lambda)$ and $\Delta(\lambda)$ for the films prepared by DC sputtering technique on a glass and PET substrates at room temperature are depicted in the figure 5 (a, b) and the figure 5 (c, d) respectively. From the figure 5 and table 4, it is clear that the two curves display the same behavior, so we can conclude from the (MSE) $\chi^2$ values that the new amorphous fitting model was appropriate in this case.

In addition, the experimental and calculated data of $\Psi(\lambda)$ and $\Delta(\lambda)$ of the NiO films on glass and PET substrates have been calculated. Fitting the experimental ellipsometric spectra of $\Psi$ and $\Delta$ as a function of wavelength allowed the calculations of the film thickness $(d)$ and the refractive and extinction coefficients $(n)$ and $(k)$ respectively of all films. The thicknesses determined by fitting the experimental ellipsometric spectra and the best-fit model parameters of $Bi_2O_3$ and NiO films are listed in Table 4.

The refractive index $(n)$ and extinction coefficients $(k)$ of $Bi_2O_3$ and NiO films on both substrates are shown in the figure 6. All films showed a similar behavior in refractive and extinction indices spectra on both substrates. The experimental and calculated values of $n$ at 550 nm of about 2.22, 2.21 and 2.58, 2.69 for $Bi_2O_3$ films on glass and PET substrates, while those of NiO films are about 2.32, 2.42 and 2.59, 2.68 on the same substrates. The extraction coefficient values at 550 nm are $4.4x10^{-2}$, $4.5x10^{-2}$ and $8.8x10^{-2}$, $9.2x10^{-2}$ for $Bi_2O_3$ films on glass and PET substrates, and $6.5x10^{-2}$, $6.9x10^{-2}$ and $6.5x10^{-2}$, $8.8x10^{-2}$ for NiO films on glass and PET substrates. As a result, $n$ and $k$ values at $\lambda_{550}$ of the $Bi_2O_3$ films decrease with those of NiO films due to an increase in the carrier concentration in the case of $Bi_2O_3$ films. For all of the NiO films, the refractive and extinction indices values at $\lambda_{550}$ is greater than those of $Bi_2O_3$ films, and a shift in the maximum value of $n$ for $Bi_2O_3$ films compare with pure NiO in the inset of the figure 6 (a and b) on glass substrate and the figure 6 (c and d) on PET substrate. The values of $n$ and $k$ of films on both substrates sharply changed in the visible region and decreases with the increasing wavelengths in the same manner.

The graphs of the experimental and calculated values of dielectric constants real $\varepsilon_r(\lambda)$ and imaginary $\varepsilon_i(\lambda)$ against wavelengths of the films are presented in the figure 7 (a, b, c and d). The values of real dielectric of films vary in a similar manner of about 4.93, 4.91 and 6.67, 7.3 at $\lambda_{550}$ for $Bi_2O_3$ films on both substrates, while the values are about 6.47, 7.1 and 0.58, 0.63 at $\lambda_{550}$ for NiO films on both substrates. The maximum values are about 9.03, 8.89 and 8.66, 9.12 for $Bi_2O_3$ films on both substrates and about 8.22, 8.67 and 8.66, 9.15 for NiO films on the same substrates before decreasing sharply with increasing wavelengths. The same behavior occurs with the imaginary dielectric of films, the maximum values of imaginary dielectric of the $Bi_2O_3$ films on both substrates of about 1.95, 2.33 and 1.63, 1.76 while the same values for NiO films are about 1.77, 1.66 and 1.94, 2.07 respectively.

The experimental and calculated graphs of the absorption coefficient $\alpha(\lambda)$ of the $Bi_2O_3$ and NiO thin films deposited on glass and PET were presented in the figure 8 (a) and (b) respectively. From this figure, the values of $\alpha$ are about $1.02 \times 10^6$ $cm^{-1}$, $1.08 \times 10^6$ $cm^{-1}$, $1.89 \times 10^6$ $cm^{-1}$ and $2.01 \times 10^6$ $cm^{-1}$ for $Bi_2O_3$ films in the visible spectra while $\alpha$ are $1.38 \times 10^5$ $cm^{-1}$, $1.48 \times 10^5$ $cm^{-1}$, $1.79 \times 10^5$ $cm^{-1}$ and $1.94 \times 10^5$ $cm^{-1}$ for NiO films. The absorption coefficient $\alpha$ in the UV range of $Bi_2O_3$ and NiO films is larger than those values in the visible and near-infrared spectra of the films on both glass and PET substrates.

By means of the absorption coefficient spectra of the films, the experimental and calculated optical gap of $Bi_2O_3$ thin films for different optical transitions modes on glass and PET substrates are shown in the figure 9 (a and b). In addition, the experimental and calculated data of the same band gap values of the NiO films on glass and PET substrates have been determined.

The experimental and calculated optical bad gap values of $Bi_2O_3$ and NiO thin films on glass and PET substrates for different optical transitions are listed in Table 5.

It is observed that, the experimental values of optical band gap of the films are in an excellent agreement with those of calculated ones. Literature results show a wide range of optical band gap values for both $Bi_2O_3$ and NiO films deposited on both substrates [26, 45-47].

### 3.3 Comparisons of optical properties and constants by means of spectrophotometry and spectro-ellipsometry

The figure 10. (a, b, c, d, e, and f) graphically illustrates the transmission, reflection and absorption spectra of $Bi_2O_3$ and NiO films deposited on glass and PET substrates using two methods: UV-vis-NIR spectrophotometry and spectro-ellipsometry in order to study and compare the optical properties and constants of the 20 nm thick $Bi_2O_3$ and NiO films in the wavelength range from 200 to 2200 nm. It is observed in this figure that the transmittance, reflectance, and absorption vary in a similar manner for all of them. The average transmission value in the visible region using UV and SE of about 84%, 82% and 79%, 76% for $Bi_2O_3$ films on glass and PET substrates, while the average transmission value is about 80%, 75% and 77%, 75% for NiO films on both substrates. The average reflectance value using UV and SE in the same range is about 10%, 14% and 14%, 18% for $Bi_2O_3$ films on glass and PET substrates respectively while it is 14%, 16% and 20.4%, 20.1% for NiO films on both substrates respectively. The average absorption value is about 5.57%, 5.39% and 12.6%, 12.5% for $Bi_2O_3$ films on glass and PET substrates respectively while it is about 9.4%, 8.9% and 7.3%, 9.8% for NiO films on glass and PET substrates respectively. However, the differences between the values of the T, R and A of the films on two substrates using two optical methods is low giving a good agreement between them.

The figure 11 demonstrates the refractive and extinction coefficients of the $Bi_2O_3$ and NiO films on glass (a and b) and (c and d) on PET substrates. It is observed from this figure that the refractive and extinction indices vary also in similar manner for all of them. The average refractive value in the visible region using UV and SE of about 2.04, 2.34 and 2.26, 2.55 for $Bi_2O_3$ films on glass and PET substrates while it is about 2.16, 2.36 and 2.65, 2.63 for NiO films on both substrates. The refractive and extinction coefficients from UV measurements are obtained from the following equation [42],

$$n = \frac{1+R}{1-R} - \sqrt{\frac{4R}{(1-R)^2} - k^2} \qquad 11$$

where R is the reflectance and $k = \frac{\alpha\lambda}{4\pi}$ is the extinction coefficient.

The figure 12 graphically illustrates the real and imaginary dielectric constants of the $Bi_2O_3$ and NiO films on glass (a) and (b) on PET substrates. It is observed from this figure that the real and imaginary values vary also in a similar manner for all of them. The average real dielectric value in the visible region using UV and SE is about 3.79, 4.45 and 4.66, 5.1 for $Bi_2O_3$ films on glass and PET substrates, while it is about 4.69, 5.55 and 7.01, 6.93 for NiO films on both substrates. The average imaginary dielectric value in the visible region using UV and SE of about 0.184, 0.24 and 0.24, 0.42 for $Bi_2O_3$ films on glass and PET substrates while it is about 0.26, 0.32 and 0.41, 0.46 for NiO films on both substrates. The real and imaginary dielectric constant curves using UV analysis are obtained from the following equation

$$\varepsilon_r = n^2 - k^2 \qquad 12$$

$$\varepsilon_i = 2nk \qquad 13$$

where $n$ and $k$ is the refractive and extinction coefficients respectively.

## 4. Conclusion

Optical characterization of the deposited $Bi_2O_3$ and NiO ultra-thin films of about 20 nm thicknesses on different types of substrates at room temperature by a DC sputtering technique have been investigated. Optical studies were done in order to calculate the transmittance, reflectance, absorption, refractive and extinction coefficients, real and imaginary dielectric constants, maximum absorption coefficient, and the optical band gap with four types of optical transitions. The optical constant spectrum in the range 200 nm - 2200 nm is recorded using spectrophotometry and spectro-ellipsometry. The estimated accuracy in the measurement of these parameters is 4%. XRD patterns of the films coated on both glass and PET substrates show an amorphous structure phase. The transmittance spectrum of these thin films shows nearly 80% to more than 87% in the visible range on both glass and PET substrates.

The optical constant values of films on both glass and PET substrates by UV method are calculated with an excellent agreement with those of SE analysis. On the other hand, excellent results obtained for the direct and indirect optical band gap determinations for different optical transition modes show that these values of $Bi_2O_3$ films range from 3.55-3 eV and from 3.4-2.9 eV on both substrates, while range from 3.75-2.9 eV and 3.5-2.7 eV for NiO films deposited on glass and PET substrates respectively. The DC sputtering technique at room temperature of high transparent $Bi_2O_3$ and NiO films on both glass and PET substrates shows that these ultra-thin films can become excellent candidates for transparent conducting oxide (TCOs) in the optoelectronic applications.

**Acknowledgement**

Authors are grateful to Nicolas Mercier, Magali Allain for providing the necessary facilities for XRD studies, Also, to Jean-Paul Gaston and Celine Eypert from Jobin Yvon Horiba Company for the spectroscopic ellipsometry measurements and to Cecile Mézière, Valerie BONNIN for help with the chemicals and corresponding equipment.

## List of Figures

Fig 1: SEM images for the deposited layers: $Bi_2O_3$, and NiO on glass and PET substrates.

Fig 2: Transmittance and Reflectance as a function of wavelength for $Bi_2O_3$ and NiO thin film on (a) glass and (b) PET substrates.

Fig 3: Absorption spectra as a function of wavelength for $Bi_2O_3$ and NiO thin film on glass and PET substrates obtained from the Lambert-Beer expression $\alpha = \frac{-lnT}{d}$ [42], where: $T$ and $d$ is the transmittance and thickness of the film (=20 nm) respectively.

Fig 4 (a and b): The optical band gap of $Bi_2O_3$ films for different optical transitions deposited on glass and (c and d) on PET substrates.

Figure 5: SE spectra of $Bi_2O_3$ and NiO films.

Fig. 6 (a and b): Experimental and calculated fitting data of the $n(\lambda)$ and $k(\lambda)$ spectra of $Bi_2O_3$ and NiO films on glass and (c and d) on PET substrates

Fig. 7 (a and b): Experimental and calculated fitting data of the $\varepsilon_r(\lambda)$ and $\varepsilon_i(\lambda)$ spectra of $Bi_2O_3$ and NiO films on glass and (c and d) on PET substrates

Fig. 8. (a): Experimental and fitting data of $\alpha(\lambda)$ curves of the $Bi_2O_3$ and NiO films on glass substrate and (b): on PET substrate.

Fig. 9 (a): Experimental and fitting data of optical band gap for different electronic transitions of $Bi_2O_3$ films on glass substrate and (b): on PET substrate.

Fig. 10 (a, c and e): UV and SE measurements of transmission, reflection and absorption spectra of $Bi_2O_3$ and NiO films on glass substrate and (b, d and f): on PET substrate.

Fig. 11 (a and b): UV and SE measurements of refractive $n$ and extinction $k$ indices of $Bi_2O_3$ and NiO films on glass (a) and (b) PET substrates.

Fig. 12 (a and b): UV and SE measurements of real and imagery of dielectric constants of $Bi_2O_3$ and NiO films on glass (a) and (b) PET substrates.

## List of Tables

Table 1: The band gap values of the $Bi_2O_3$ and NiO films deposited on glass and PET substrates by UV-vis-NIR spectrophotometer for different optical transitions modes.

Table 2: Show physical parameters of glass substrate by new amorphous model.

Table 3: Show physical parameters of PET substrate by Tauc-Lorentz model.

Table 4: Parameters of the New-amorphous fitting method for $Bi_2O_3$ and NiO films deposited on glass and PET substrates.

Table 5: The experimental and calculated values of optical band gap by spectro-ellipsometry.

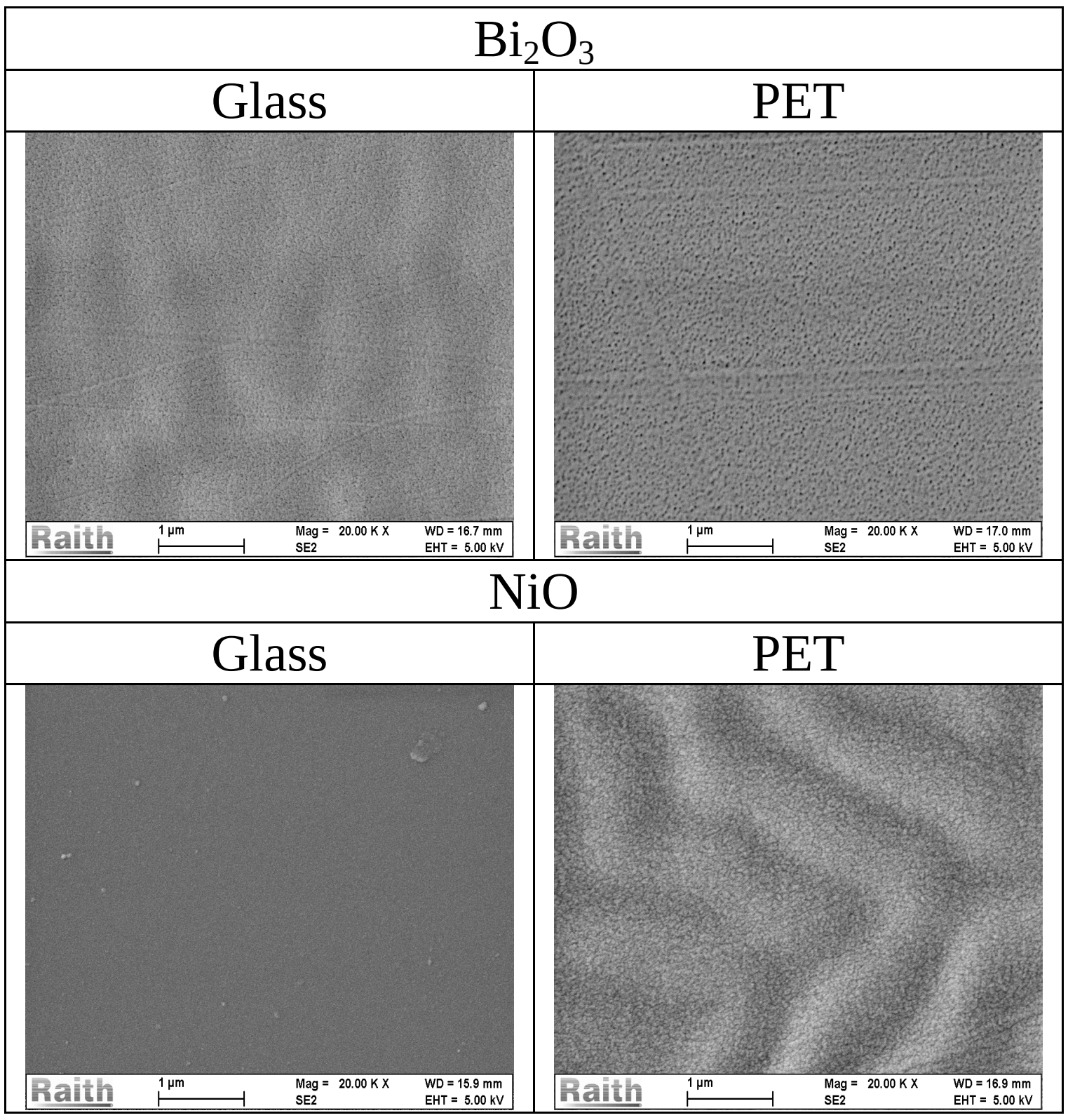

Fig. 1

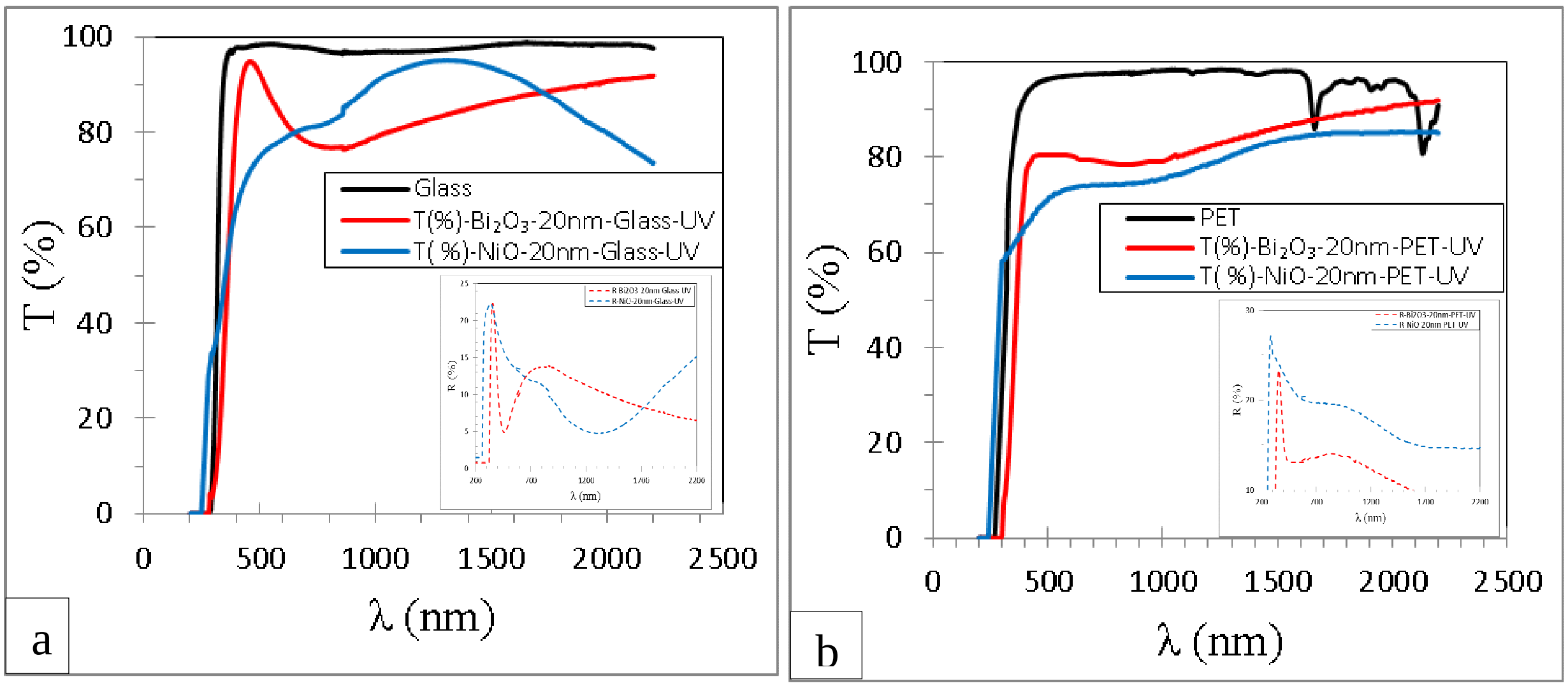

Fig. 2

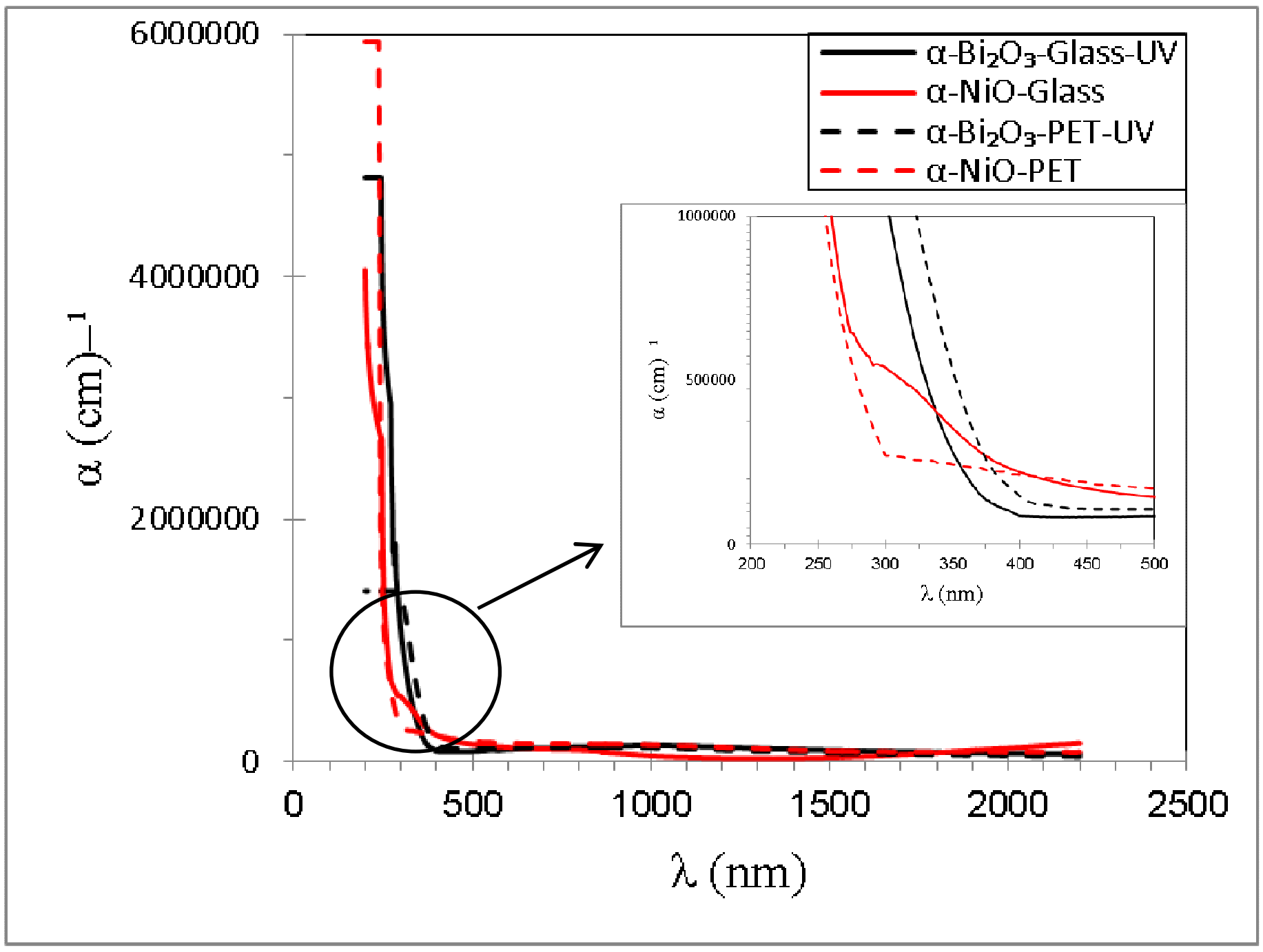

Fig. 3

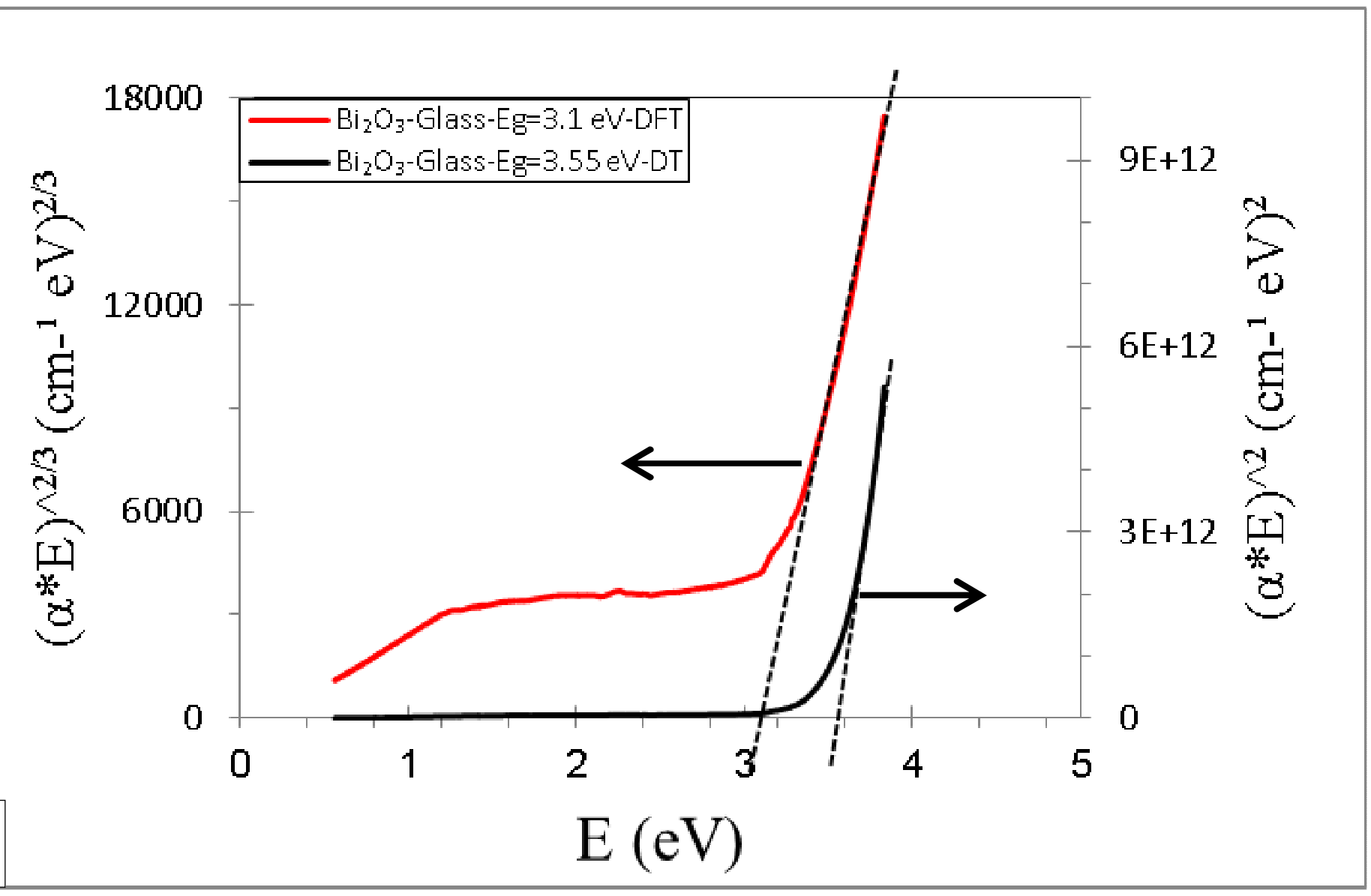

a

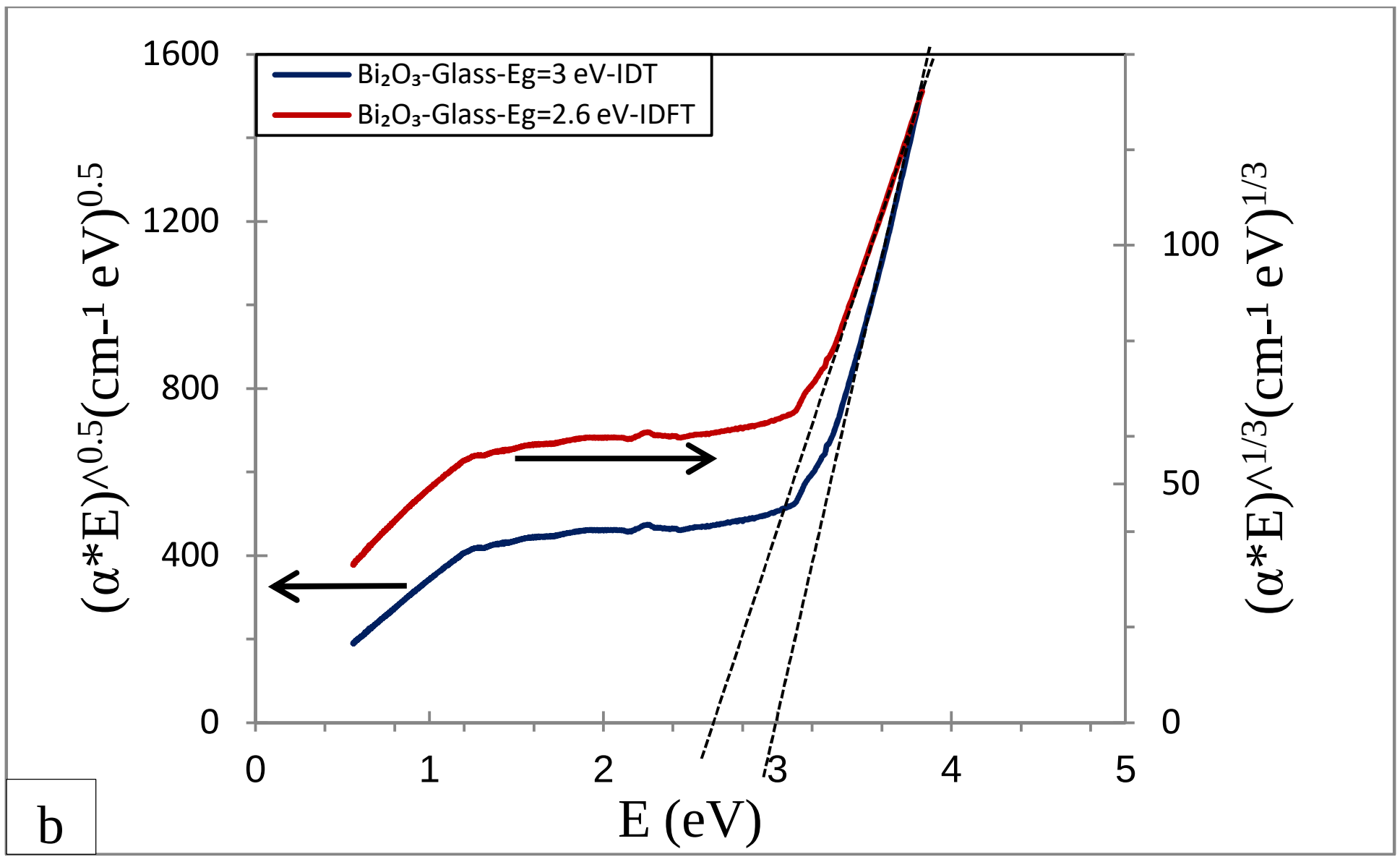

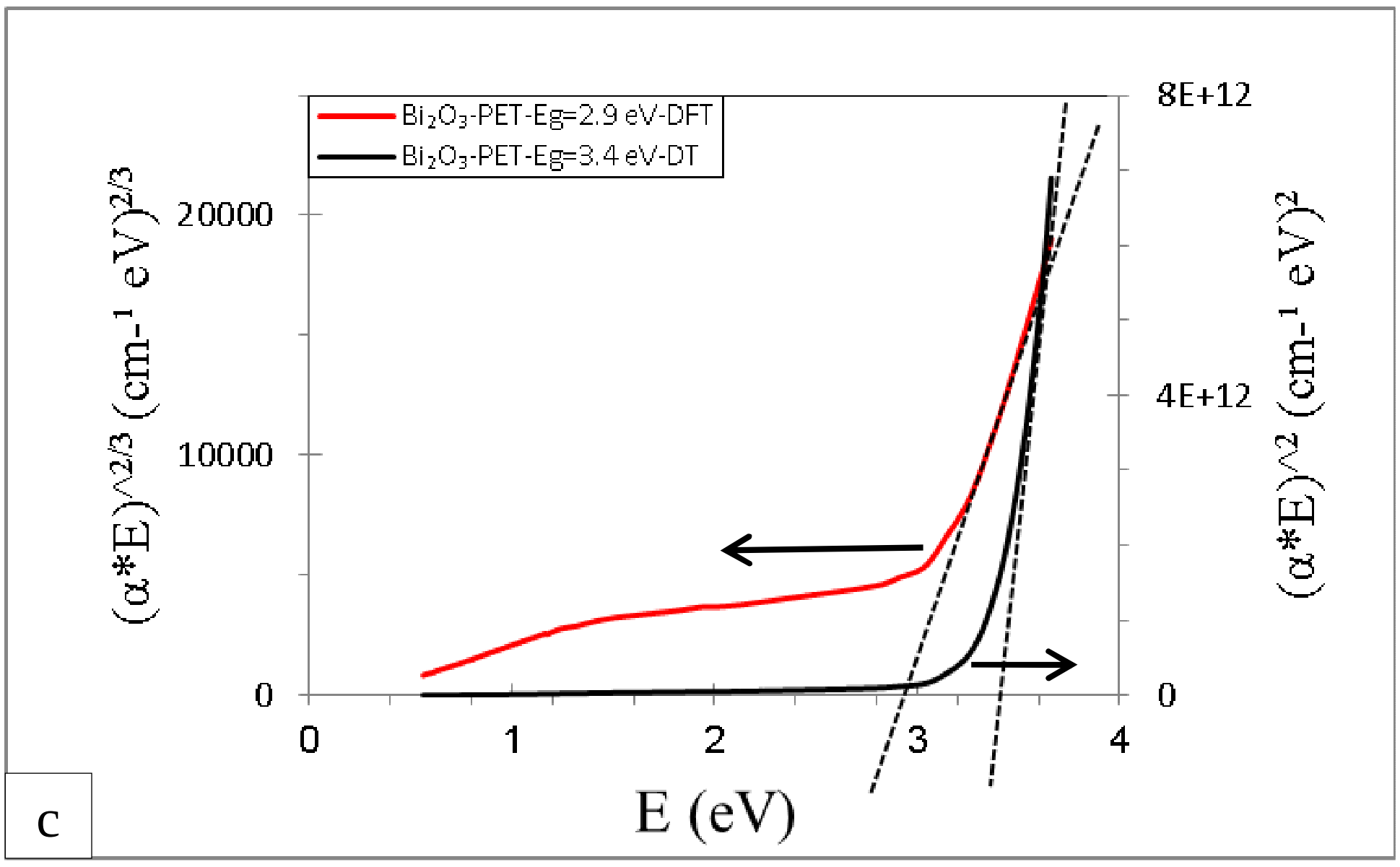

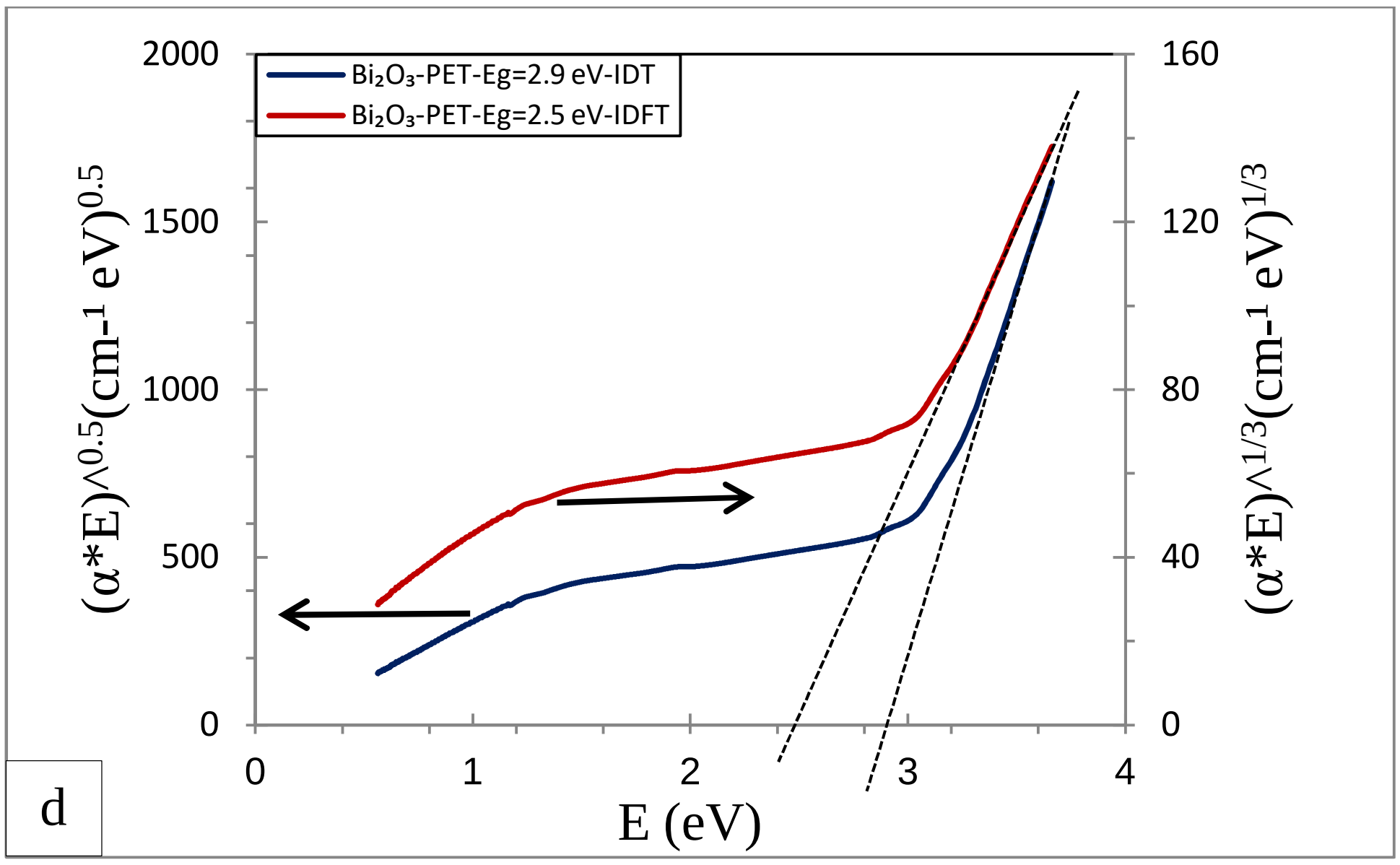

Fig. 4

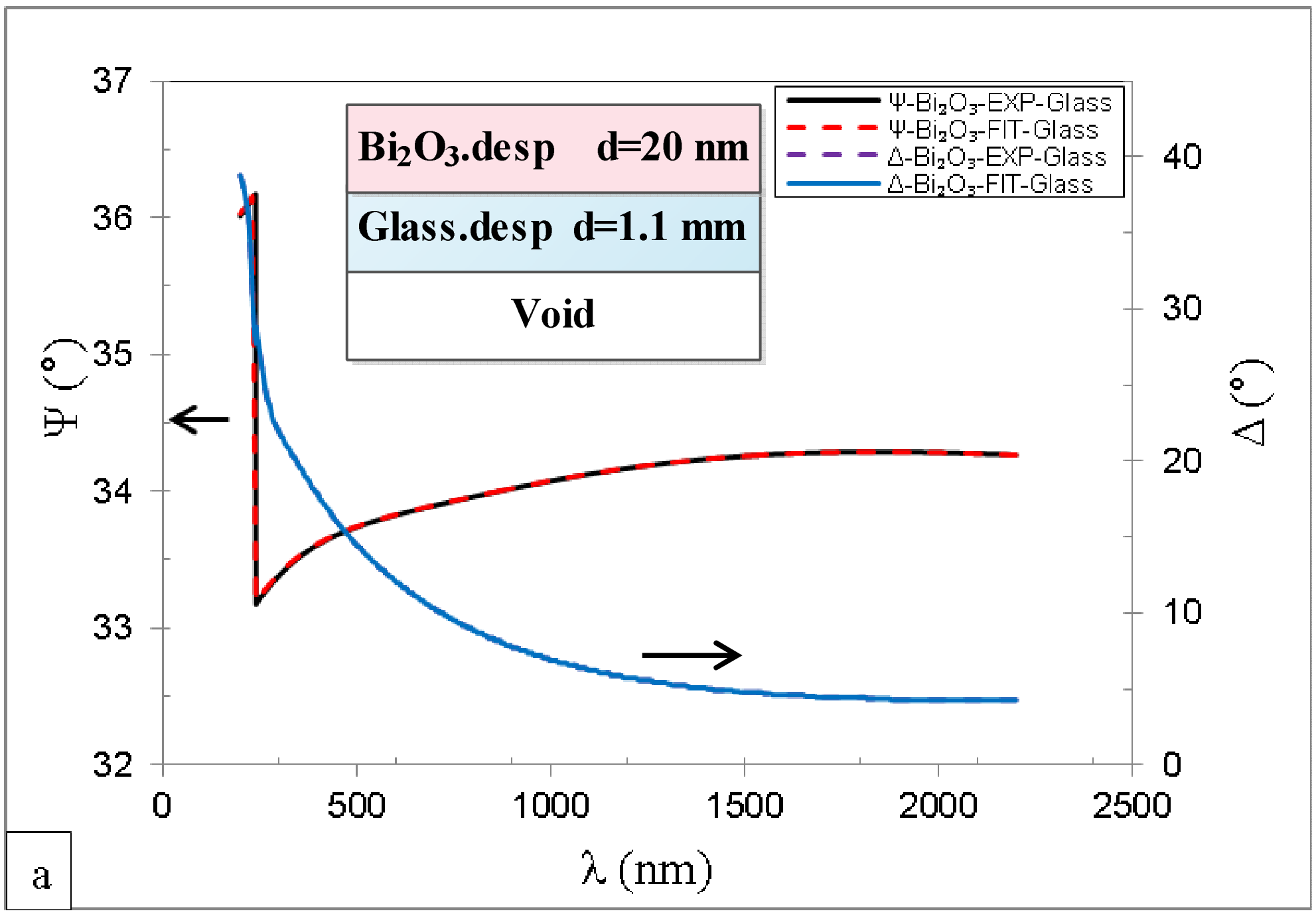

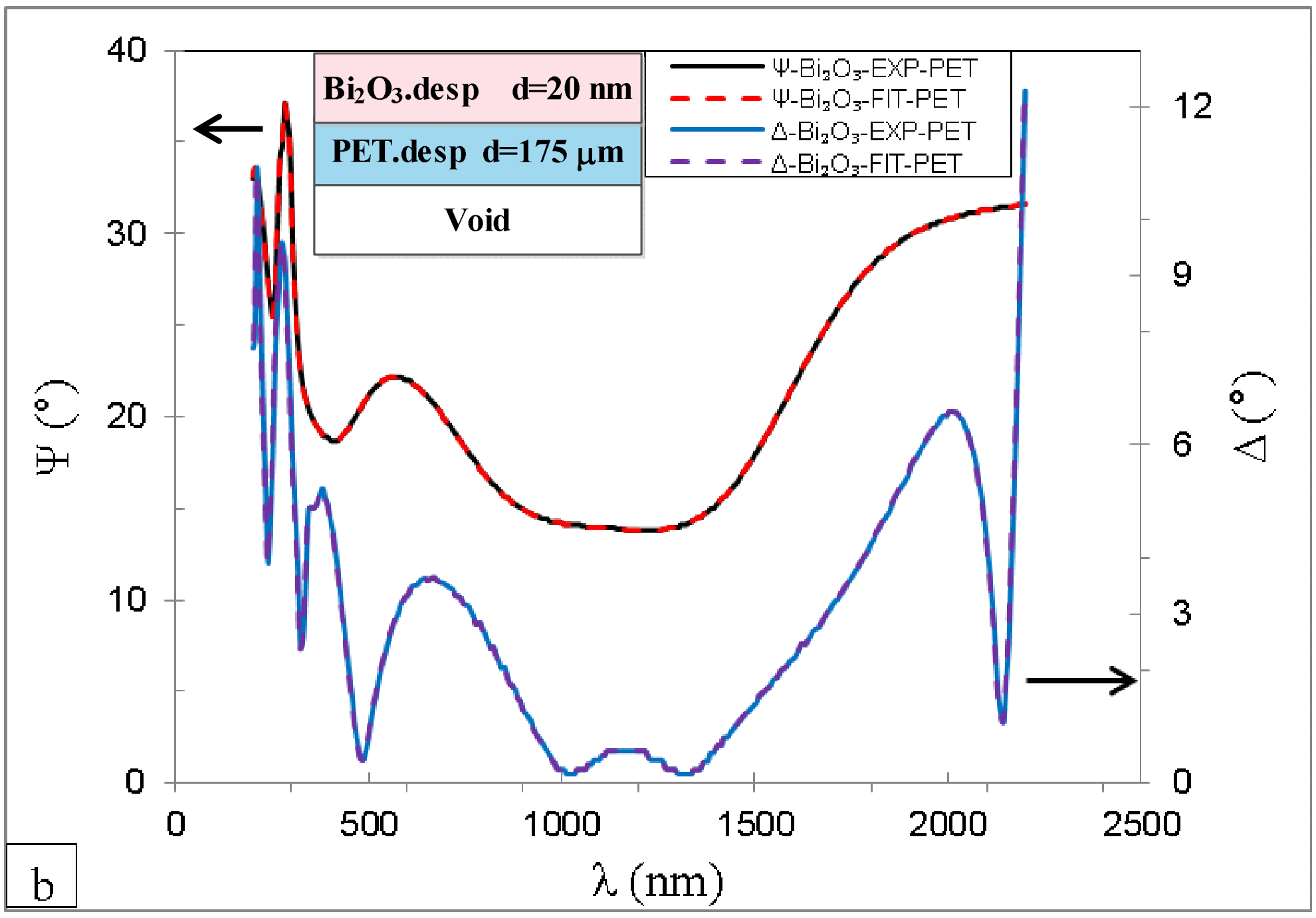

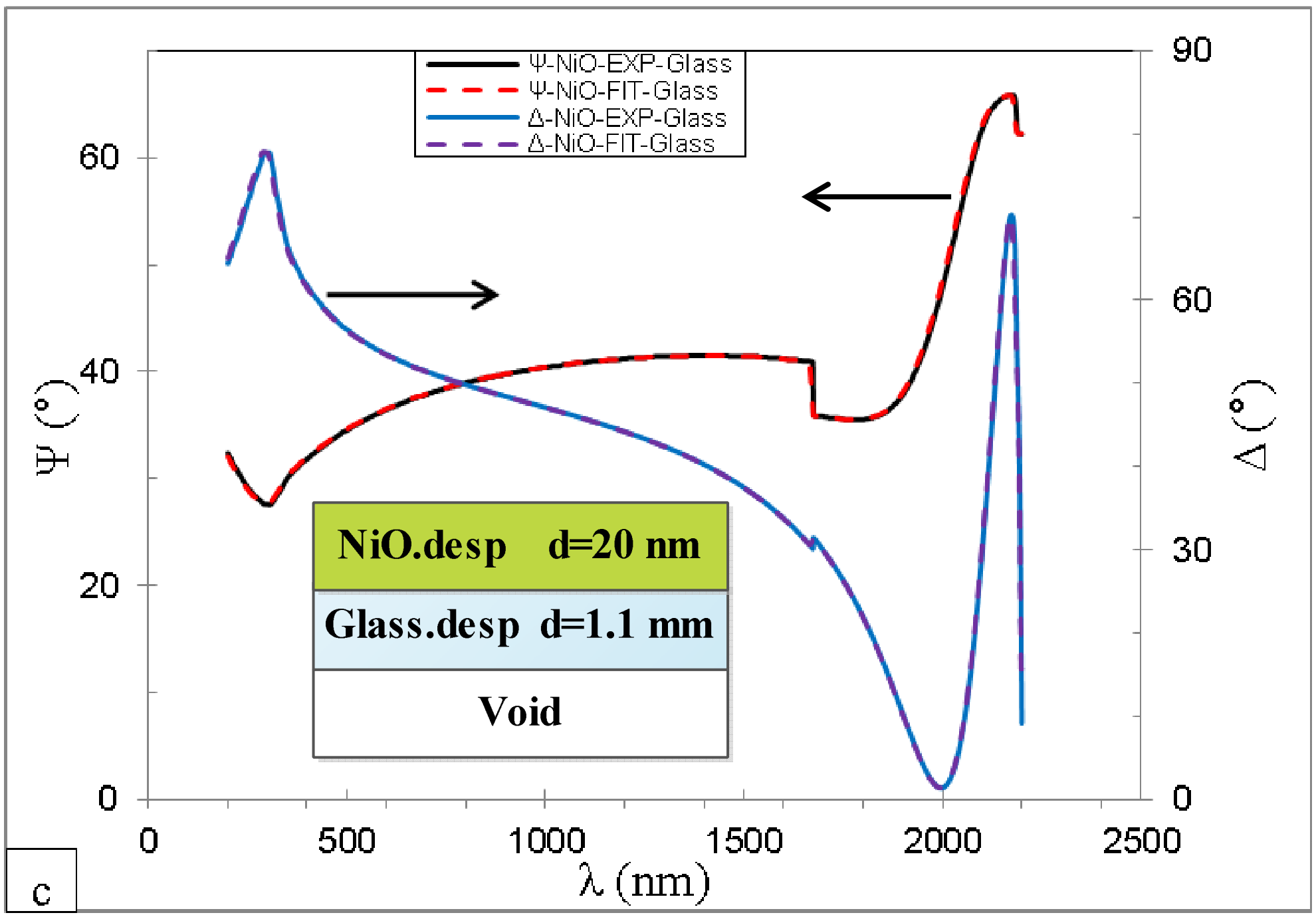

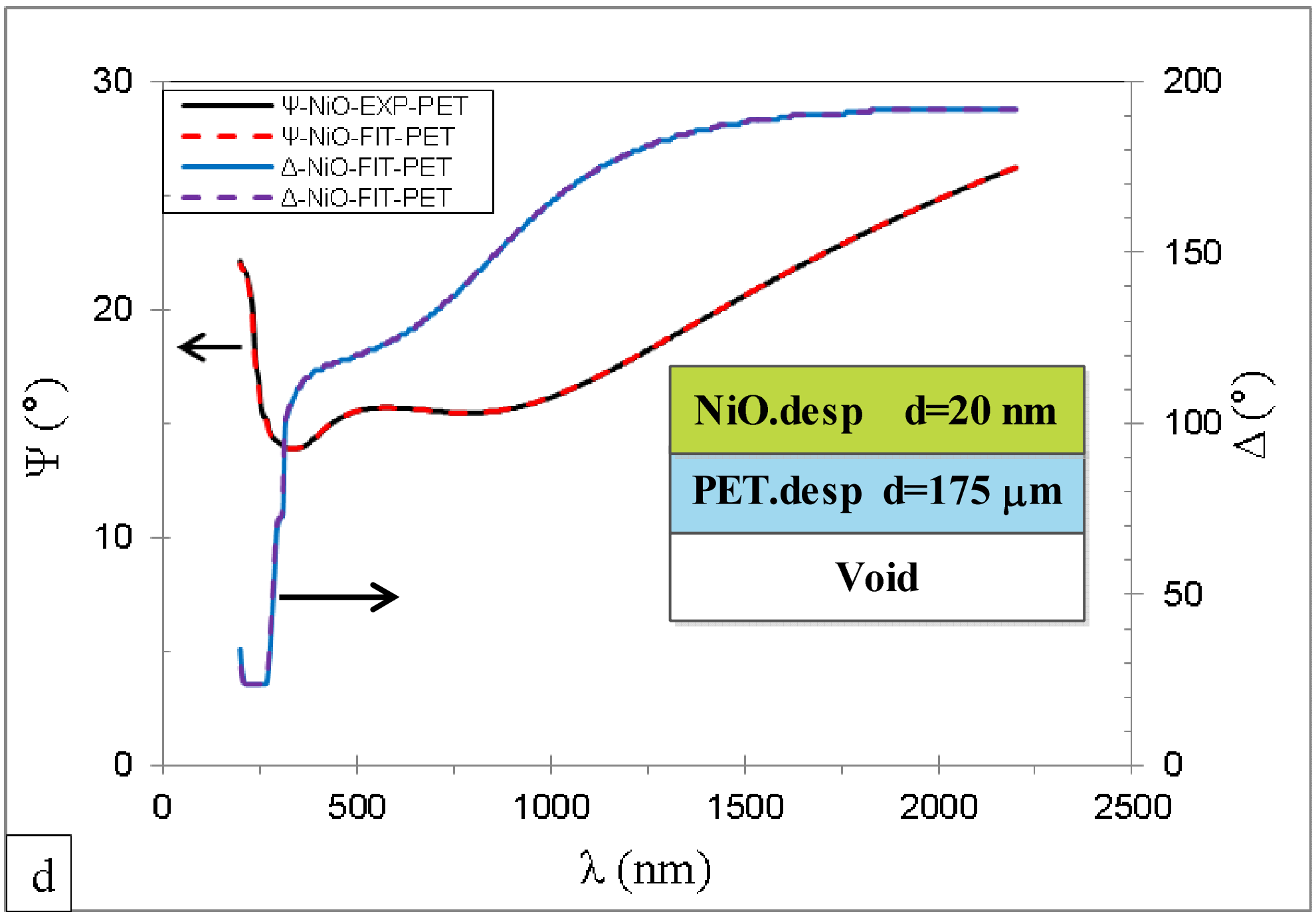

Fig. 5

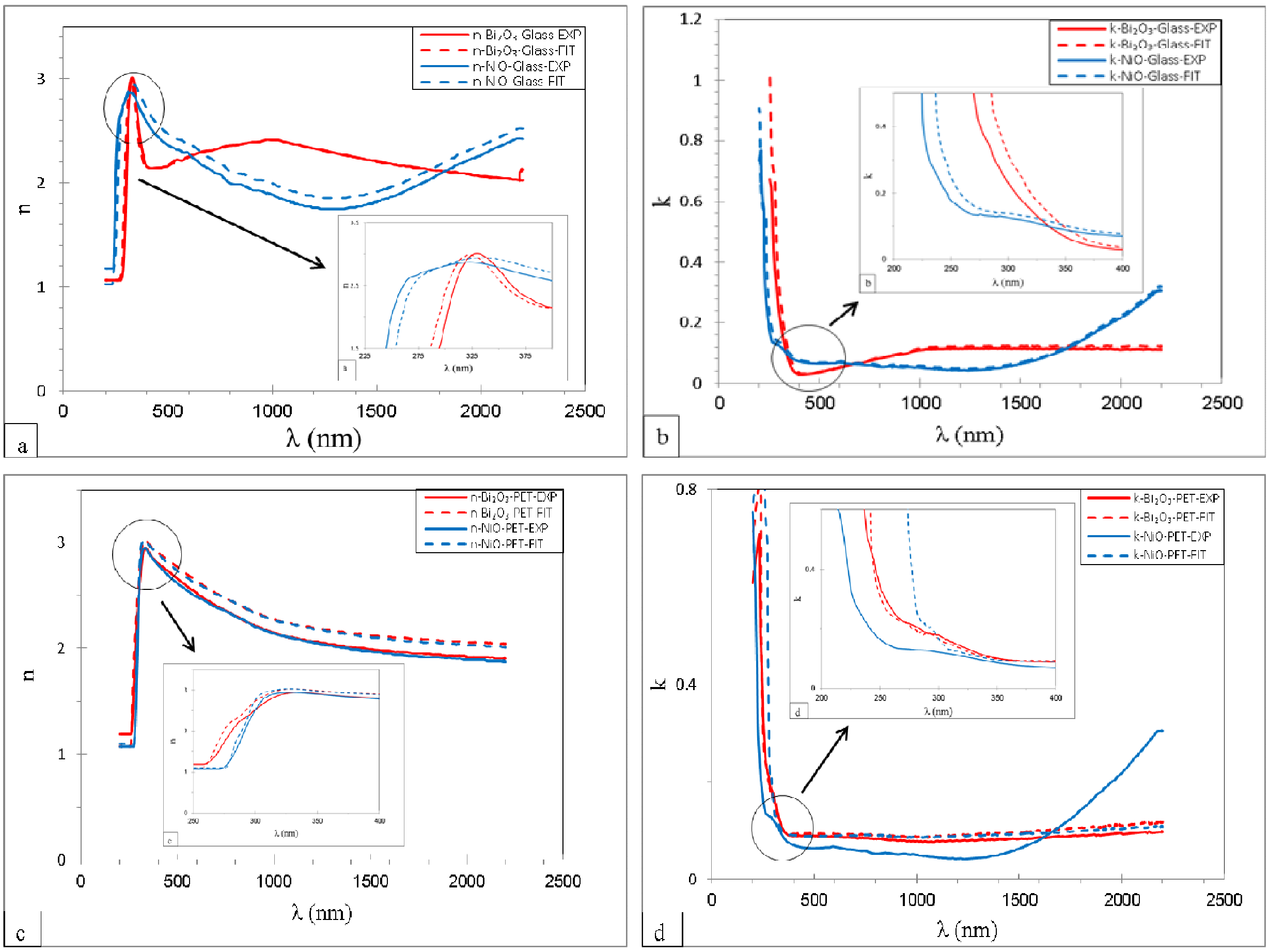

Fig. 6

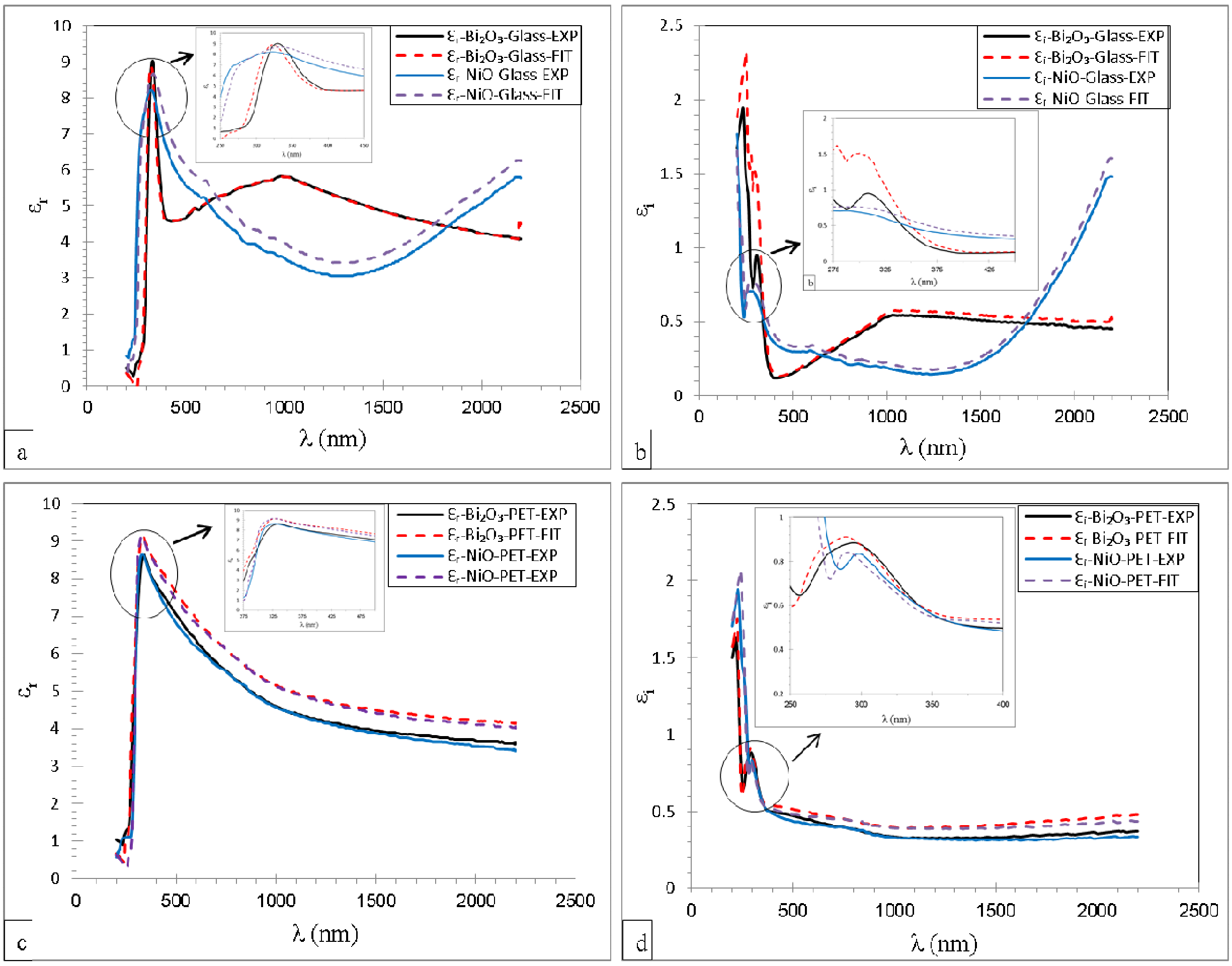

Fig. 7

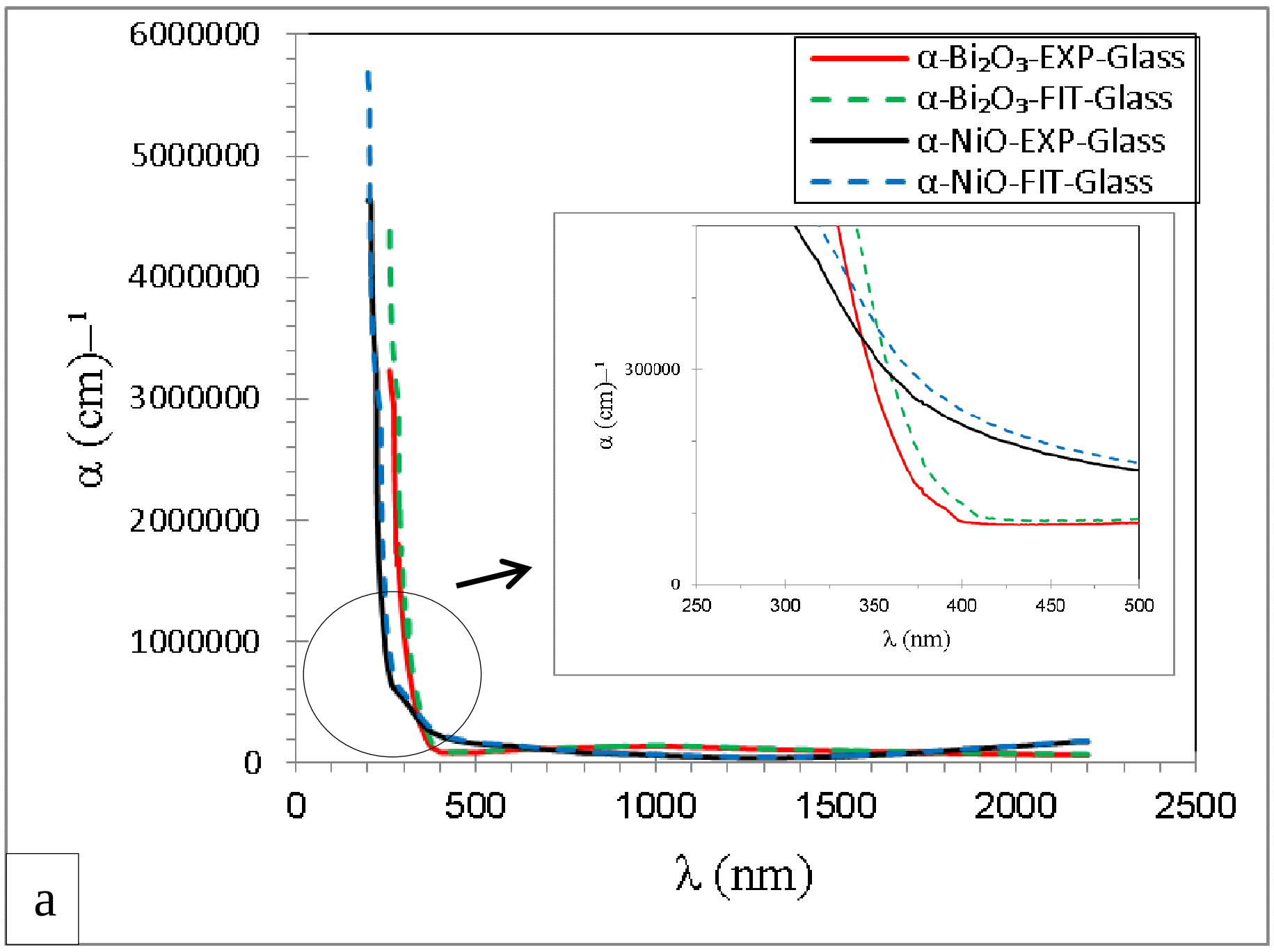

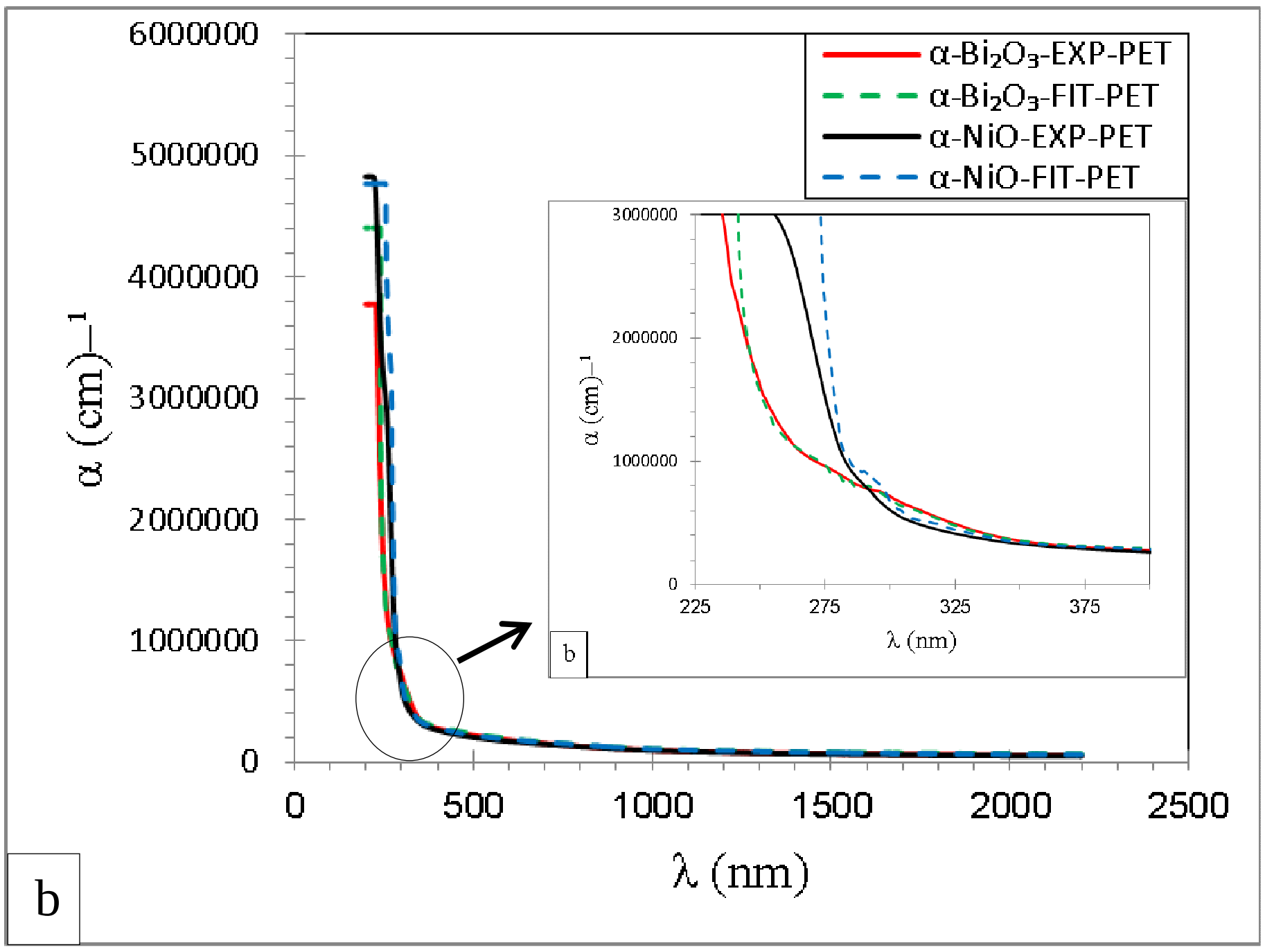

Fig. 8

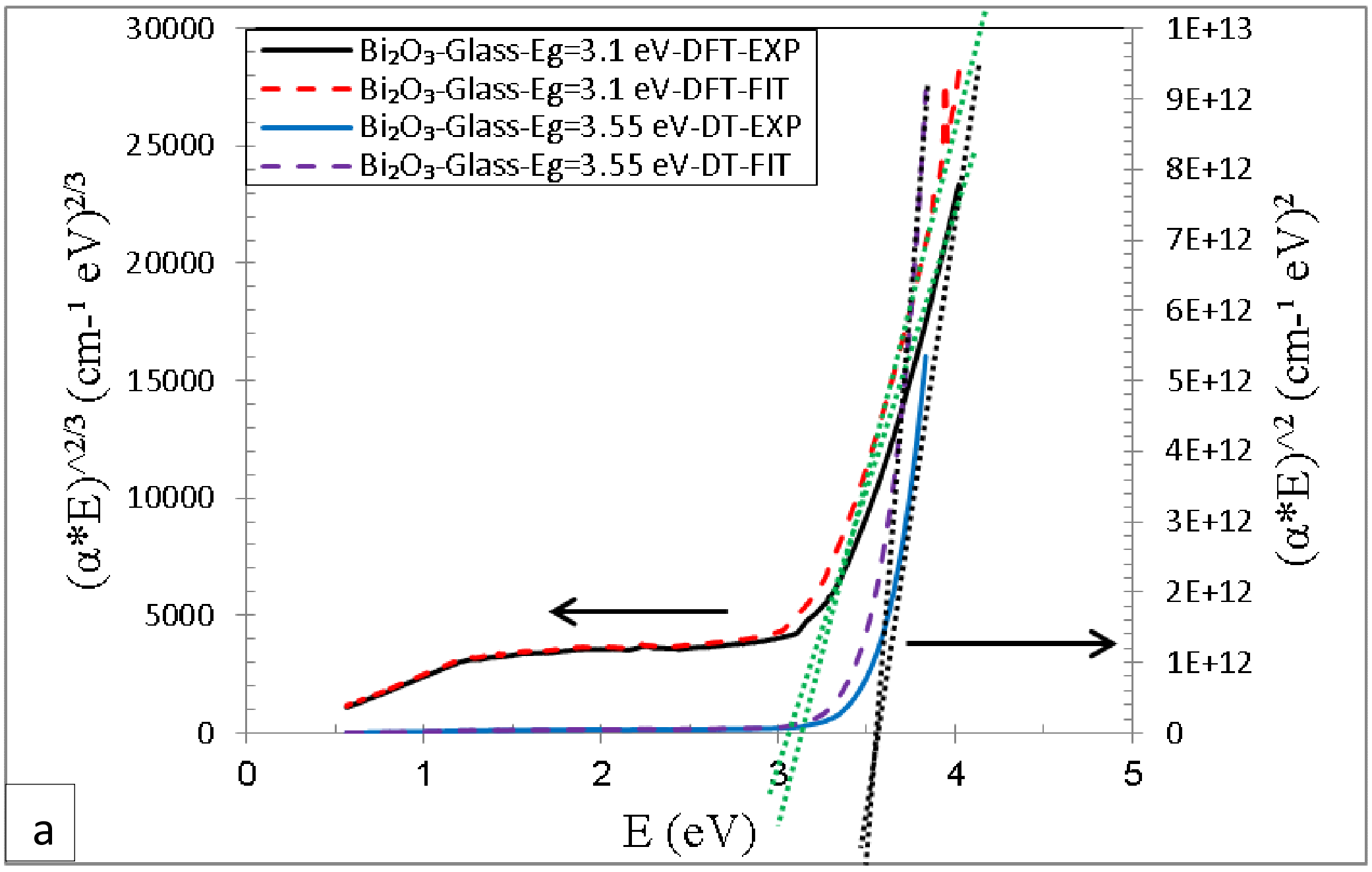

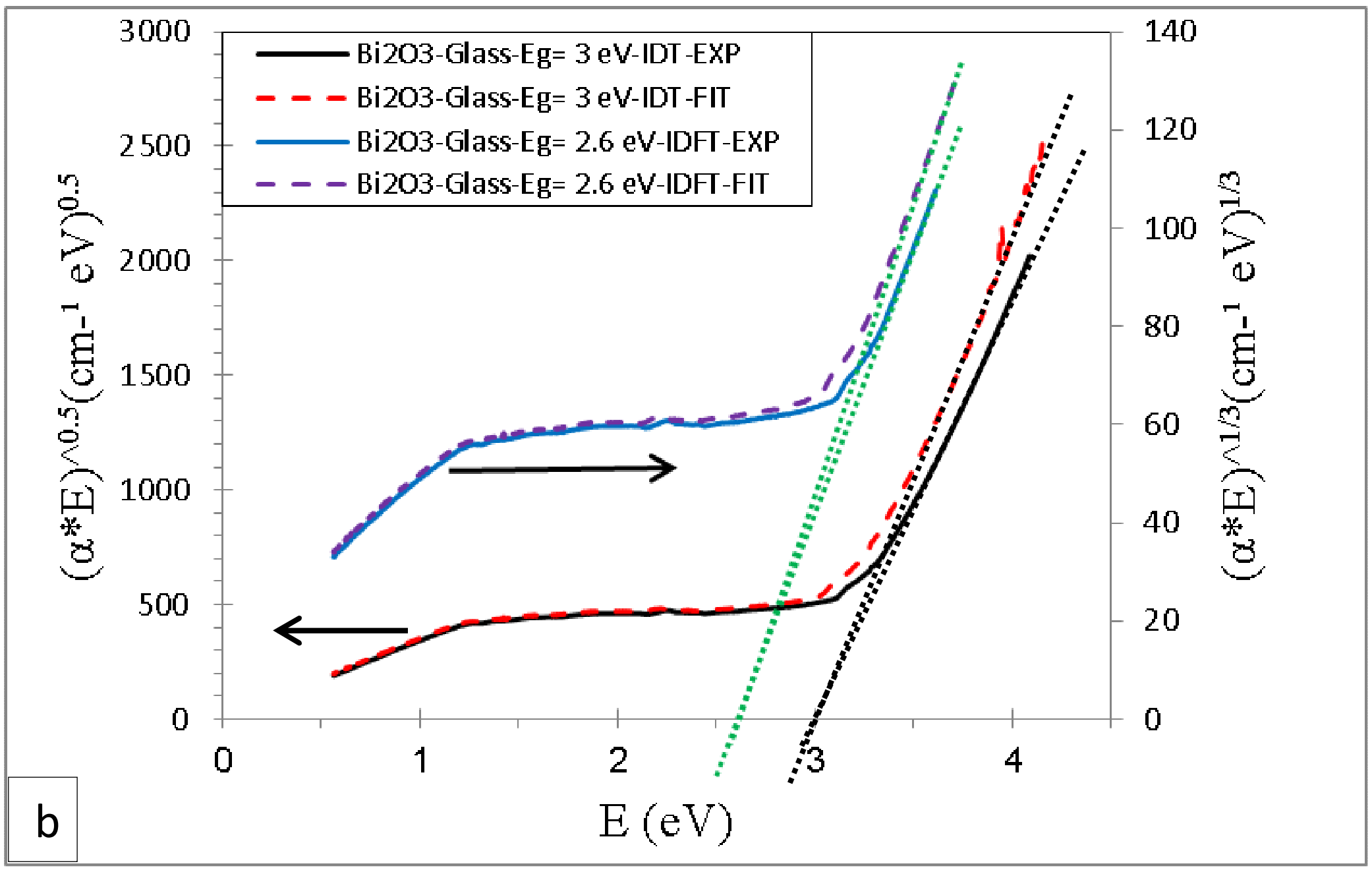

Fig. 9

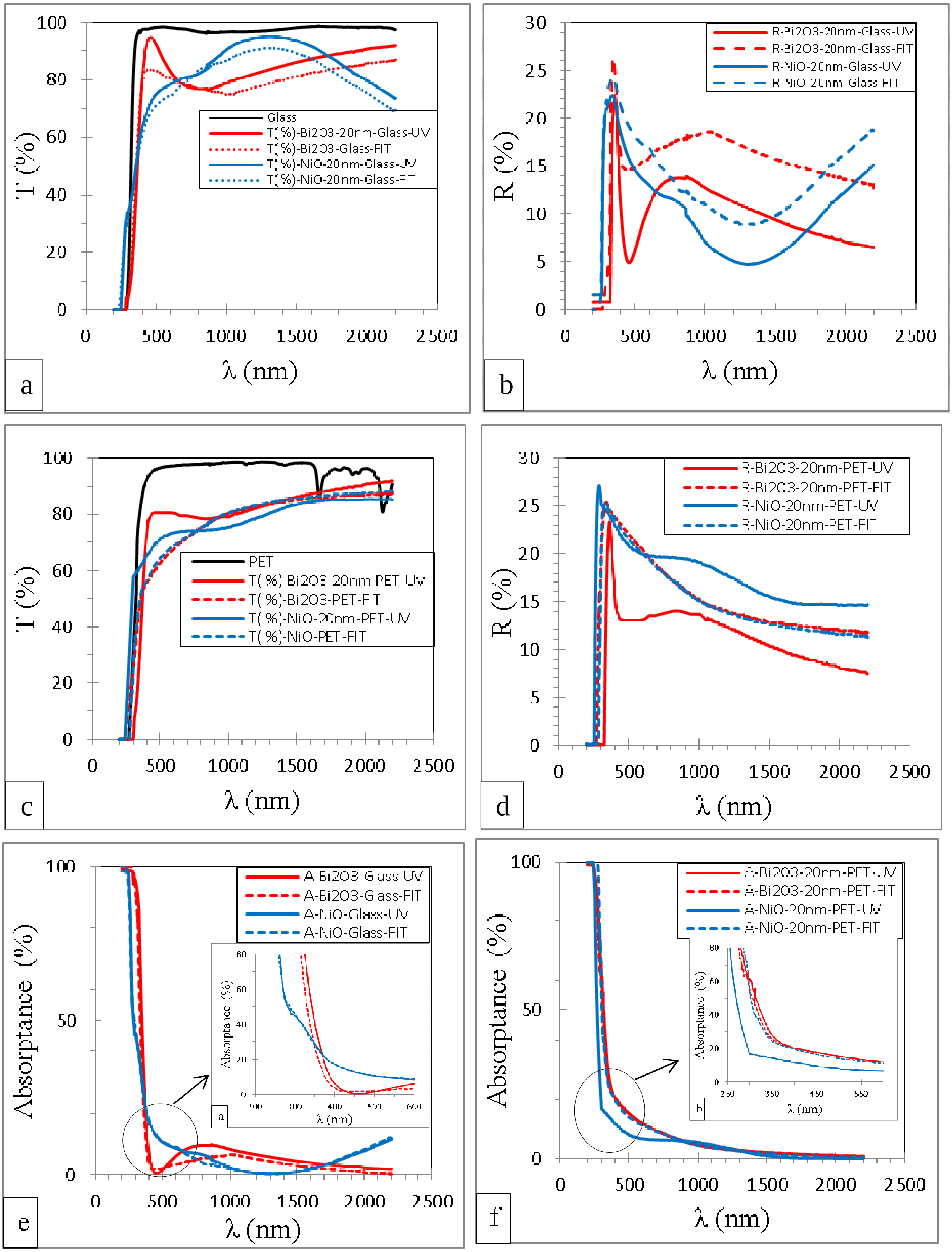

Fig. 10

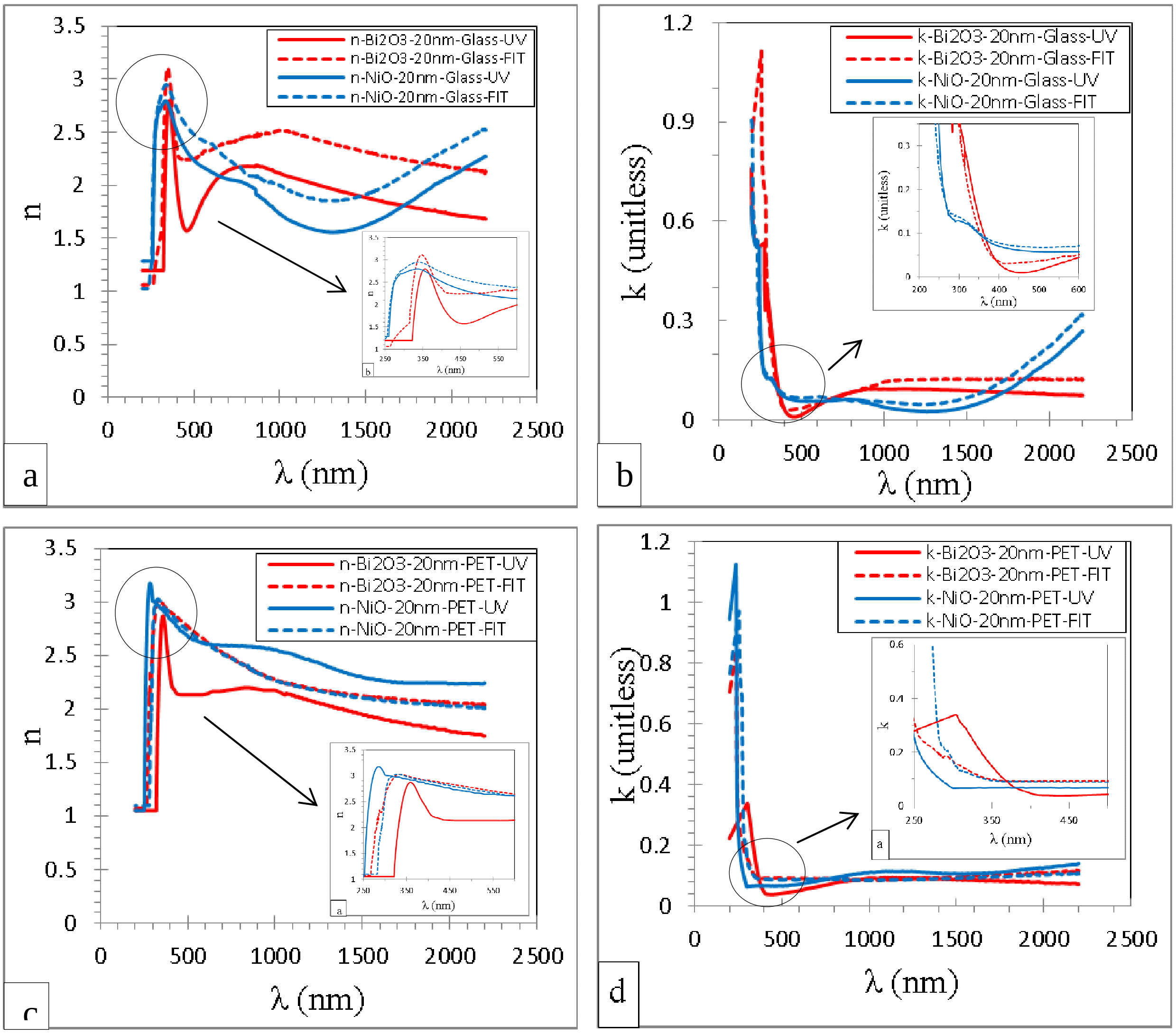

Fig. 11

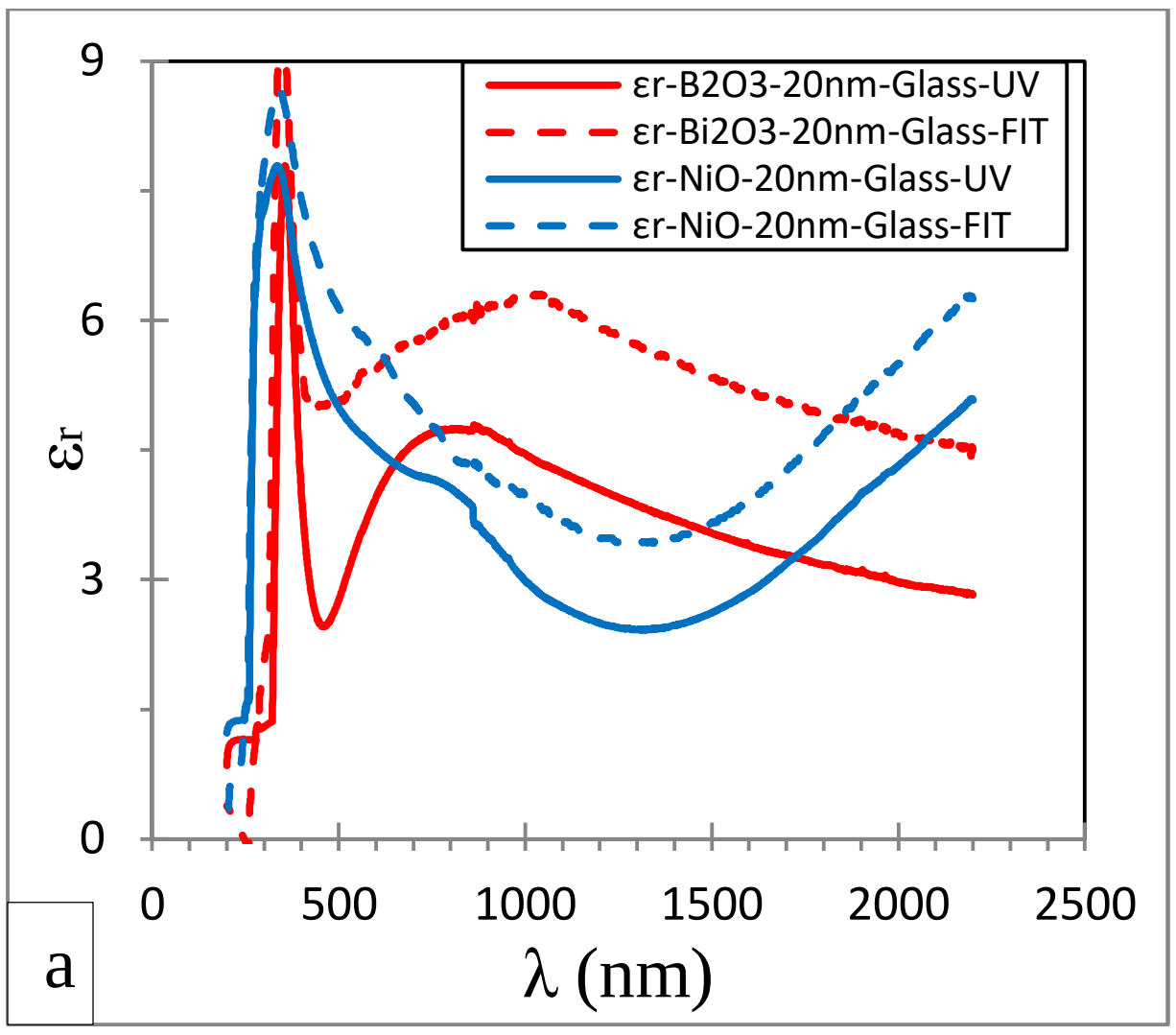

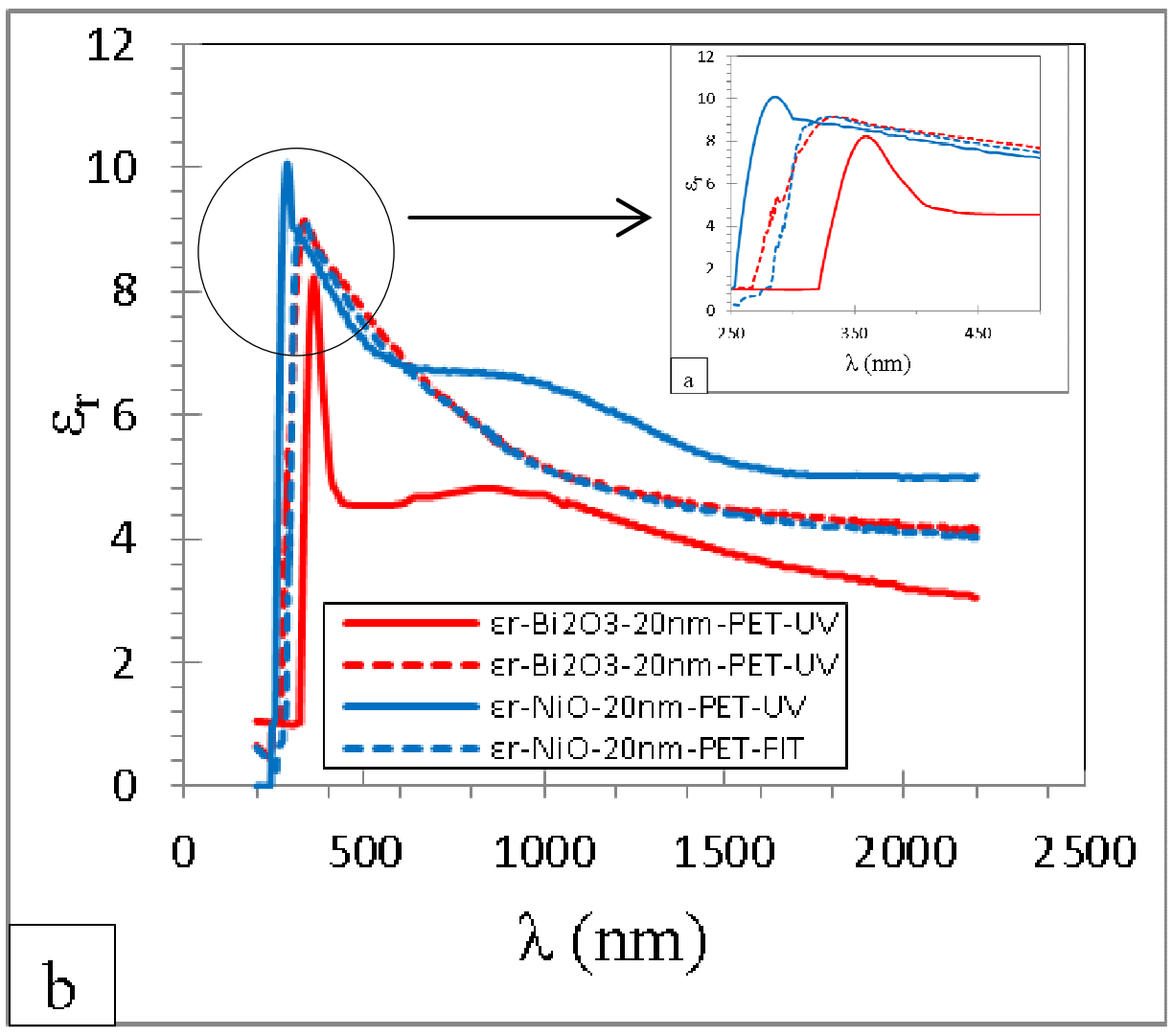

Fig. 12

| | Thickness (nm) | Spectrophotometry | | | | | | | |
|---|---|---|---|---|---|---|---|---|---|
| Substrate | | Glass | | | | PET | | | |
| Optical transition | | DT | DFT | IDT | IDFT | DT | DFT | IDT | IDFT |
| $Bi_2O_3$ | 20 | 3.55 | 3.1 | 3 | 2.6 | 3.4 | 2.9 | 2.9 | 2.5 |
| NiO | 20 | 3.75 | 2.7 | 2.9 | 2.5 | 3.5 | 2.4 | 2.7 | 2.45 |

Table 1

| **Model parameters** | **Glass substrate** |
|---|---|
| $\boldsymbol{n_\infty}$ | 0.63 |
| $\boldsymbol{\omega_g}$ | 3.23 |
| $\boldsymbol{f_j}$ | 0.05 |
| $\boldsymbol{\omega_j}$ | 41.77 |
| $\boldsymbol{\Gamma_j}$ | 2.31 |
| $\boldsymbol{AOI}$ | 70 |
| $\boldsymbol{d}$ (nm) | 1100000 |

Table 2

| **Model parameters** | **PET substrate** |
|---|---|
| $\boldsymbol{\varepsilon_\infty}$ | 2.51 |
| $\boldsymbol{E_g}$ | 3.68 |
| $A_1$ | 6.39 |
| $E_1$ | 4.17 |
| $C_1$ | 0.17 |
| $A_2$ | 22.17 |
| $E_2$ | 4.95 |
| $C_2$ | 0.58 |
| $A_3$ | 8.69 |
| $E_3$ | 6.19 |
| $C_3$ | 0.18 |
| $\boldsymbol{AOI}$ | 70 |
| $\boldsymbol{d}$ (nm) | 175000 |

Table 3

| **Model parameters** | **Glass** | | **PET** | |
|---|---|---|---|---|
| | **$Bi_2O_3$** | **NiO** | **$Bi_2O_3$** | **NiO** |
| $\chi^2$ **(MSE)** | 2.66 | 0.07 | 0.12 | 0.014 |
| $\boldsymbol{n_\infty}$ | 14.70±1.69 | 14.92±0.04 | 10.81±0.03 | 0.356±0.0006 |
| $\boldsymbol{\omega_g}$ | 3.76±0.05 | 0.53±0.01 | 3.23±0.01 | 6.37±0.01 |
| $\boldsymbol{f_j}$ | 1.74±8.22 | 18.41±0.05 | 14.52±0.05 | 0.02±0.01 |
| $\boldsymbol{\omega_j}$ | 2.226±0.724 | 0.49±0.001 | 1.27±0.001 | 1.208±0.001 |
| $\boldsymbol{\Gamma_j}$ | 8.265±0.724 | 0.006±0.001 | 0.66±0.01 | 2.76±0.01 |
| $\boldsymbol{AOI}$ | 70.967±2.725 | 70.584±0.001 | 71.754±0.001 | 72.408±0.023 |
| $\boldsymbol{d}$ **(nm)** | 20.0±0.8 | 20.01±0.01 | 20.21±0.02 | 20.44±0.01 |

Table 4

| Optical band gap determinations by spectro-ellipsometry | | | | | | | | | |
|---|---|---|---|---|---|---|---|---|---|
| Materials | Thickness (nm) | Glass substrate | | | | | | | |
| | | DT | | DFT | | IDT | | IDFT | |
| | | $E_{g_{EXP}}$ | $E_{g_{FIT}}$ | $E_{g_{EXP}}$ | $E_{g_{FIT}}$ | $E_{g_{EX}}$ | $E_{g_{FIT}}$ | $E_{g_{EXP}}$ | $E_{g_{FIT}}$ |
| $Bi_2O_3$ | | 3.55 | 3.55 | 3.1 | 3.1 | 3 | 3 | 2.6 | 2.6 |
| NiO | | 3.75 | 3.75 | 2.7 | 2.7 | 2.9 | 2.9 | 2.5 | 2.5 |
| Materials | Thickness (nm) | PET substrate | | | | | | | |
| | | $E_{g_{EXP}}$ | $E_{g_{FIT}}$ | $E_{g_{EXP}}$ | $E_{g_{FIT}}$ | $E_{g_{EX}}$ | $E_{g_{FIT}}$ | $E_{g_{EXP}}$ | $E_{g_{FIT}}$ |
| $Bi_2O_3$ | | 3.4 | 3.4 | 2.9 | 2.9 | 2.9 | 2.9 | 2.5 | 2.5 |
| NiO | | 3.5 | 3.5 | 2.4 | 2.4 | 2.7 | 2.7 | 2.45 | 2.45 |

Table 5